\documentclass[%
 aip,
 jmp,%
 amsmath,amssymb,
 reprint,%
]{revtex4-1}

\usepackage{graphicx}
\usepackage{dcolumn}
\usepackage{bm}

\usepackage{chngpage}

\newcommand{\vc}{\bm{v_c}}
\newcommand{\pc}{\bm{p_c}}

\newcommand{\uc}{u_c}
\newcommand{\mc}{n_c}
\newcommand{\Vc}{V_c}
\newcommand{\ekinc}{e_{c}^{\text{kin,s}}}
\newcommand{\ekinct}{e_{c}^{\text{kin, t}}}

\newcommand{\rzi}{\bm{r}_i}

\newcommand{\pzi}{\bm{p}_i}

\newcommand{\kb}{k_{\text{B}}}

\newcommand{\xc}{\bm{x}_c}

\newcommand{\xe}{{\bm{x}}_e}
\newcommand{\xtot}{{\bm{x}}_{\text{tot}}}
\newcommand{\txtot}{{\tilde{\bm{x}}}_{\text{tot}}}

\newcommand{\lambdac}{\bm{\lambda}_c}

\newcommand{\stot}{S_{\text{tot}}}
\newcommand{\tstot}{{S}}
\newcommand{\etot}{E_{\text{tot}}}
\newcommand{\Vtot}{V_{\text{tot}}}
\newcommand{\mtot}{N_{\text{tot}}}
\newcommand{\ptot}{\bm{P}_{\text{tot}}}

\newcommand{\Venv}{V_{\text{env}}}
\newcommand{\menv}{N_{\text{env}}}
\newcommand{\uenv}{U_{\text{env}}}
\newcommand{\penv}{\bm{p}_{\text{env}}}
\newcommand{\senv}{S_{\text{env}}}
\newcommand{\xenv}{\bm{x}_{\text{env}}}
\newcommand{\venv}{\bm{v}_{\text{env}}} 

\newcommand{\len}{\left \langle}
\newcommand{\ren}{\right \rangle}

\newcommand{\thetam}{\theta_{\text{micro}}}
\newcommand{\rhosat}{\rho_{\text{sat}}}
\newcommand{\psat}{p_{\text{sat}}}

\newcommand{\Pcal}{\mathcal{P}}
\newcommand{\Qcal}{\mathcal{Q}}
\newcommand{\Lcal}{i\mathcal{L}}

\begin{document}

\preprint{AIP/123-QED}

\title[Systematic Coarse-Graining in Nucleation Theory]{Systematic Coarse-Graining in Nucleation Theory}

\author{M. Schweizer}
 \email{marco.schweizer@math.ethz.ch}
 \affiliation {ETH Zurich, Department of Materials, Polymer Physics, Vladimir-Prelog-Weg 5, 8093 Zurich, Switzerland.}
\author{L. M. C. Sagis}%
 \email{leonard.sagis@wur.nl}
\affiliation{
Food Physics Group, Wageningen University, Bomenweg 2, 6703 HD Wageningen, The Netherlands and ETH Zurich, Department of Materials, Polymer Physics, Vladimir-Prelog-Weg 5, 8093 Zurich, Switzerland.
}%

\date{\today}

\begin{abstract} 
In this work we show that the standard method to obtain nucleation rate-predictions with the aid of atomistic Monte-Carlo simulations leads to nucleation rate predictions that deviate $3-5$ orders of magnitude from the recent brute-force molecular dynamics simulations [J. Diemand, R. Ang\'{e}lil, K. K. Tanaka, and H. Tanaka, J. Chem. Phys. \textbf{139}, 074309 (2013)] conducted in the experimental accessible supersaturation regime for Lennard-Jones argon. We argue that this is due to the truncated state space literature mostly relies on, where the number of atoms in a nucleus is considered the only relevant order parameter. We here formulate the nonequilibrium statistical mechanics of nucleation in an extended state space, where the internal energy and momentum of the nuclei is additionally incorporated. We show that the extended model explains the lack in agreement between the molecular dynamics simulations by Diemand et al.\ and the truncated state space. We demonstrate additional benefits of using the extended state space; in particular, the definition of a nucleus temperature arrises very naturally and can be shown without further approximation to obey the fluctuation law of McGraw and Laviolette. In addition, we illustrate that our theory conveniently allows to extend existing theories to richer sets of order parameters.
\end{abstract}

\pacs{Valid PACS appear here}
\maketitle

\begin{quotation}
\section{Introduction}

Nucleation is the initial step of a phase transformation, a process ubiquitous from the smallest to the largest physical length-scales: the fabrication of new nanoscale material for various future applications such as the design of new microchips requires a control on the morphology of nucleated material\cite{Yang:2006, Longuet:2014}. On the micrometer scale, nucleation is for instance relevant in the development of new drugs; here, different nucleation conditions can lead to different packing arrangements of the molecular substance, which would be pharmaceutically classified as different drugs. Therefore a precise handle on the nucleation process is deeply required\cite{Horst:2002aa}. On the higher end of length scales, research has been conducted\cite{Kuba:2010} to study cloud seeding - the injection of a substance into clouds to induce rainfall - offering the potential to control the weather and the atmosphere. On cosmic length-scales nucleation leads to creation of interstellar dust after super novae. The dust modifies the spectrum of interstellar radiation, and its composition and time-scale of 
formation are of key interest\cite{KrugelB2003}. In all these phenomena, small nuclei have to overcome an activation barrier to initiate the phase-transformation and to grow deterministically. The rate at which small nuclei are capable to escape from the initial fluctuation dominated growth regime is the fundamental quantity of interest in nucleation theory. This nucleation rate has been predicted through a variety of methods ranging from brute force molecular dynamics (MD) simulations\cite{Diemand:2013,Chkonia:2009,Kraska:2006,Tanaka:2005,Tanaka:2011,Wedekind:2007,Yasuoka:1998} to phenomenological, semi-phenomelological and statistical mechanics approaches \cite{KalikmanovB2013, Girshick:1990,Dillmann:1991,Delale:1991, Kalikmanov:1995, Laaksonen:1994, Reguera:2004} among others.

MD simulates all involved atoms in the nucleation process by solving their equations of motion and can be used to obtain the exact nucleation rates under various initial conditions of the metastable phase. Since MD conducts computer experiments with systems of nano sized dimensions, up to limits of a few micrometers, these studies favour regimes in which the nucleation rate is large, so that despite of the microscopic system size critically sized nuclei pop up frequently enough and adequate statistics on the obtained data can be ensured. In the prototypical studied nucleation of liquid drops in a metastable gas, MD works in the large supersaturation regime where the metastable phase is very dense. This is in strong contrast to real experiments. Most laboratory experiments\cite{Schmitt:1984, Wright:1993, Viisanen:1993, Anisimov:2001} are performed at low supersaturations and measure nucleation rates less than  $10^{10} \text{cm}^{-3} s^{-1}$. For the nucleation of argon the latest experimental setups through the supersonic Nozzle (SSN) experiment\cite{Sinha:2001} achieved to study nucleation rates up to $10^{17} \text{cm}^{-3} s^{-1}$. Only recently, in $2013$, large-scale MD simulations\cite{Diemand:2013} were capable to operate in the experimentally accessible regime set by the SSN. In Fig.~\ref{Fig_Comp_efficiency} we have reported the studied nucleation rate in MD simulations of argon in some of the literature throughout the past $20$ years - unless sophisticated large-scale simulations are exploited, MD would typically study nucleation regimes that differ by $5-10$ orders of magnitude from the SSN experiment and $12-17$ orders of magnitude from typical experiments.
\begin{figure}[!htb]
\centering
\includegraphics[width=8.7cm]{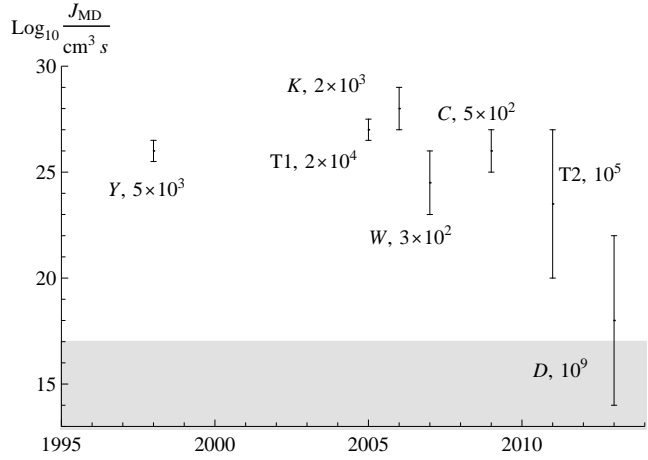}
\caption{Illustration of accessible nucleation rate $J_{\text{MD}}$ in selected molecular dynamics studies\cite{Diemand:2013,Chkonia:2009,Kraska:2006,Tanaka:2005,Tanaka:2011,Wedekind:2007,Yasuoka:1998} of argon from $1997-2013$. We marked the first letter of the family name of the first author of the study together with the number of used atoms. The shaded area shows the experimental accessible regime.  }
\label{Fig_Comp_efficiency}
\end{figure} 
MD cannot be used to study realistic experimental regimes unless tremendeous computational effort is made, which is the main drawback of the technique. Coarse-grained models allow to bridge the time- and length scale  gap between atomistic dynamics and nucleation as perceived on a coarser scale. Such models offer the possibility to obtain predictions in the experimental accessible nucleation regime and offer hence an appealing alternative to brute-force MD.
 
The development of coarse-grained models to capture the relevant macroscale physics of nucleation received tremendeous amount of interest. These studies mainly focus on obtaining predictions for the shape and hight of the nucleation landscape $V_S(\mc)$ that prohibits small nuclei to reach the critical size $\mc^{*}$. The barrier height $\Delta V_S$ of $V_S$ is particularily important and enters the nucleation rate through an exponential factor $e^{-\Delta V_S}$. Relatively minor errors in $\Delta V_S$ are therefore known to give rise to lack in predictions of the nucleation rate of orders of magnitude. In the literature $\kb T V_S(\mc)$ is known as the Gibbs free energy of formation of a nucleus of size $\mc$ in a metastable phase at temperatur $T$ and pressure $P$. A long history of phenomenological, semi-phenomenological and statistical mechanics models have been obtained to predict the work of formation. In particular, the classical nucleation theory\cite{KalikmanovB2013}, the internally consistent classical nucleation theory\cite{Girshick:1990}, the semiphenomenological model (SP) \cite{Dillmann:1991,Delale:1991, Kalikmanov:1995}, the revised SP model \cite{Laaksonen:1994}, the mean-field kinetic theory (MFKT)\cite{Kalikmanov:2006,KalikmanovB2013}, the extended modified liquid drop dynamic nucleation thoery \cite{Reguera:2004} and many more. In our work we aim to obtain quantitative predictions of $V_S$ from first principle atomistic Monte-Carlo (MC) simulations thereby avoiding the often crude approximations that mesoscopic theories exploit to obtain the nucleation landscape $V_S$. 

Standard MC simulations\cite{Kusaka:1998b, Oh:1999, Oh:2000,Chen:2001,Yoo:2001,Gonzalez:2014,Wolde:1998} in the literature exploit small atomistic systems with $N$ atoms, pressure $P$ and temperature $T$ to sample the probability $\phi_S(\mc)$ of occurence of a nucleus of size $\mc$. The choice of the isobaric-isothermal ensemble in computer simulations mimics the conditions of many real experiments which control exactly these variables to induce the nucleation process. The link of the microscopically obtained probability $\phi_S(\mc)$ with the macro-physics manifests itself in the relationship $V_S(\mc) = - \log(\phi_S(\mc))$. The probability $\phi_S(\mc)$ is vanishingly small around the critical size $\mc^{*}$, but the MC simulations can be biased through importance sampling techniques such as umbrella-sampling to access regions of the atomistic phase space that contain a nucleus of size $\mc$. MC simulations obtain  $\phi_S(\mc)$ even in the experimentally relevant nucleation regime thereby offering a tool that is potentially superior to brute-force MD. In order to obtain nucleation rate predictions $J$, $\phi_S$ must however enter an adequate evolution equation obtained through supplementary dynamic considerations of the macroscale nucleation process. The Zeldovich equation\cite{KalikmanovB2013}
\begin{equation}
\begin{split} 
& \frac{\partial f(\mc)}{\partial t} =  \frac{\partial}{\partial \mc}  D_{\mc,\mc}  \left[ \frac{\partial f(\mc)}{\partial \mc} + f(\mc) \frac{\partial V_S(\mc)}{\partial \mc} \right]
\label{Zeldovic_eq}
\end{split}
\end{equation} 
recognized as thermodynamically admissible\cite{Schweizer:2014} is probably the most popular to capture the dynamics of the distribution function $f(\mc)$ of a nucleus of interest. Here, $f(\mc)$ denotes the probability that the considered nucleus has size $\mc$ and $D_{\mc,\mc}$ is the diffusion coefficient of the diffusion process described by Eq.~(\ref{Zeldovic_eq}). The Zeldovich equation can be solved approximately in the stationary limit to obtain the nucleation rate as $J=J_0 e^{-\Delta V_S}$, where the prefactor $J_0$ is related to kinetic coefficients and the barrier shape\cite{KalikmanovB2013}.
 
We will illustrate that when the nucleation rate of argon droplets in a supersaturated argon gas is obtained through the standard formula $J=J_0 e^{-\Delta V_S}$, where $\Delta V_S$ is constructed from the aforementioned MC technique, a nucleation rate is predicted that deviates $3-5$ orders of magnitude from exact brute-force MD predictions by Diemand and coworkers\cite{Diemand:2013}. This is particularily severe since the simulations conducted by Diemand et al.\ consider the practically relevant supersaturation regime. Furthermore, the argon particles are described by the Lennard-Jones model and this is one of the most basic and often used microscopic models to study the nucleation process. The standard approach to obtain the nucleation barrier by MC simulations is therefore incomplete.  

We argue that this is an artefact of relying on the fully truncated state space $\xc^{\text{trc}} = (\mc)$ to describe the nuclei degree of freedom on the coarse-grained scale. Our attempt to improve nucleation rate predictions obtained from MC simulations has its fundament in $1966$ in the classical work\cite{Feder:1966} by Feder et al.\ who already noticed the need to extend the aforementioned truncated state space in order to account for certain dynamic effects that  cannot be captured by the use of a single state variable. In particular, Feder et al.\ realized that in the nucleation process a significant amount of latent heat is released indicating that the nuclei exhibit rather strong energy fluctuations on top of their size fluctuations. They showed that a proper inclusion of the internal energy $\uc$ into the nuclei state space leads to corrections to the nucleation rate of $1-3$ orders of magnitude. The nucleation landscape in their work is therefore extended to two dimensions such that $V_S = V_S(\mc, \uc)$. This is confirmed by other work\cite{Barrett:2008,Barrett:1994,Wyslouzil:1992,Schweizer:2014} focusing on nonisothermal corrections. Another argument in favour of equipping the state space with additional degrees of freedom traces back to the ``translational-rotational paradox" initiated by Lothe and Pound around $1962$\cite{Lothe:1962}. They realized that nuclei are not static objects, but can move and rotate within the metastable phase. In their semi-quantum mechanical treatment they tried to incorporate the translational and rotational contributions into $V_S(\mc)$ without extending the nuclei state space and ended up with corrections of $10^{17}$ to the nucleation rate. After incorporating their correction, theoretical predictions of the nucleation rate deviated from experimentally obtained rates so outreageously that the term ``translational-rotational paradox" was coined. Many attempts\cite{Reiss:1998,Reiss:1997,Ruth:1988,Reiss:1967b,Reiss:1967a,Reguera:2001q} tried to resolve the paradox. In this respect the work\cite{Reguera:2001q} of Reguera and Rubi presents an elegant solution by incorporating additional degrees of freedom that characterize the nuclei. In particular, they argued that a proper inclusion of the momentum $\pc$ and angular momentum $\bm{l}_c$ in the state space of the nucleus leads to much smaller corrections of the order of $10^2-10^5$ for typical nucleation regimes. The landscape would hence be extended to $V_S = V_S(\mc, \pc, \bm{l}_c)$.

The idea of extending the state space to capture relevant effects is not new and for certain order parameters, the extended nucleation landscape $V_S$ has even been studied by statistical mechanics\cite{Schaaf:1999oparam, Kusaka:2003oparam, Kusaka:1999oparam}. But in this paper we aim to lay the basis for the nonequilibrium statistical mechanics of nucleation with extended state spaces. We here restrict the study to the specific choice $\xc = (\mc, \pc, \uc)$ and to homogeneous nucleation in a one-component system. We develop a coarse-graining procedure that is reliably predicting nucleation rates in the low nucleation rate regime where critically sized nuclei are sufficiently large so that their dynamics can be captured in the continuum limit through the use of Fokker-Planck equations. This Fokker-Planck equation involves diffusion in $\xc$-space and describes the dynamics of $f(\xc)$, the probability that at a certain time a specific nucleus has property $\xc$. Our coarse-graining procedure relies on the GENERIC (General Equation for Non-Equilibrium Reversible-Irreversible Coupling) projection-operator technique\cite{HCO:POT,HCO} and yields atomistic expressions for the building blocks of the aforementioned Fokker-Planck equation. In particular, we obtain a fundamental expression for the nucleation landscape $V_S(\xc)$ that can easily be sampled in MC simulations and we show that the diffusion coefficients in the Fokker-Planck equation are a simple generalization of expressions already available\cite{Lundrigan:2009, Auer:2004} for the fully truncated case $\xc^{\text{trc}} = (\mc) $. Through our nonequilibrium statistical mechanics we obtain contributions to the nucleation barrier that have already been found through mesoscopic irreversible thermodynamics arguments concerning momentum corrections and by McGraw and Laviolette for energy fluctuations. In particular, our framework allows to rigorously define a nucleus temperature $T_c$ that obeyes the non-Gaussian behavior derived by McGraw and Laviolette\cite{McGraw:1995} which has been confirmed in direct MD simulations of nucleation\cite{Wedekind:2007aa,Diemand:2014ads}. We apply our approach to the most recent MD simulation predictions\cite{Diemand:2013} of Diemand et al.\ and show that the lack of prediction of $3-5$ orders of magnitude by the truncated theory can be eliminated. On the example of MFKT which uses the truncated state space we illustrate further that we can conveniently extend existing theories to richer state spaces thereby improving the quality of their predictions.

In Sec.~\ref{Sec_CG} we develope the coarse-graining scheme for nucleation with extended state spaces from microscopic arguments. We show that the nucleation landscape $V_S$ is deeply related to the system entropy and that $V_S$ is the fundamental static building block generating the nucleation dynamics on the coarse-grained scale. We discuss the nucleation landscape and the evolution equation governing $f(\xc)$ in some detail and relate it to existing literature results. In Sec.~\ref{sec_MFKT} we apply our results to extend the state space of the MFKT of nucleation and discuss the nucleation of argon droplets in a metastable gas. Finally, in Sec.~\ref{sec_LJNucleation} we use our coarse-graining procedure to obtain the nucleation rate in metastable Lennard-Jones gas. We show that the standard coarse-graining procedure leads to erronoreous estimates as compared to brute-force MD simulations\cite{Diemand:2013} and that these can be corrected for by our theory.

\end{quotation}

\section{Detailed Derivation of Coarse-Graining Procedure}
\label{Sec_CG} 
 
\subsection{Relevant State Variables}
\label{Sec_statevar}

We aim to investigate nucleation in a metastable one component system. Prominent examples are nucleation of liquid drops in a gas phase or nucleation of gas bubbles in a liquid phase. We characterize the state $\xc = (\mc, \pc, \uc)$ of the nuclei by means of their number of atoms $\mc$, their total momentum $\pc$ and their internal energy $\uc$. The incorporation of the total momentum allows for Brownian motion of the nuclei and the total energy as an additional variable is necessary to treat nonisothermal effects. The dynamics of a single nucleus is captured by the distribution function $f_t(\xc)$ describing the probability that the nucleus has degrees of freedom $\xc$ at time $t$; the time-argument will subsequently be suppressed in the notation. The nuclei are assumed to evolve independently, so that the evolution of each nucleus does not affect other nuclei and the single-nucleus distribution function $f(\xc)$ is sufficient to capture the relevant macro-scale physics. We hence neglect the break-up of a nucleus into several nuclei or coalescence effects. We further characterize the full system in which nucleation takes place by its total energy $\etot$, total number of particles $\mtot$, total momentum $\ptot$ and volume $\Vtot$ so that its state is captured by $\xtot = (\mtot, \ptot, \etot, \Vtot)$. Clearly, as $\xtot$ consists solely of conserved quantities, the state vector $\xtot$ remains constant during the entire evolution. The state variables $\xc$ (or respectively $f(\xc)$) and $\xtot$ are merely a splitting of degrees of freedom to capture the essential characteristics of the nucleus and its environment. However, $\xtot$ is not the state of the nucleus ambient phase, but involves both, the nucleus and ambient phase degrees of freedom. In Sec.~\ref{SEC_ST_VAR} we show that upon a suitable transformation, the variables $\xtot$ and $\xc$ can indeed be used to split the degrees of freedom into nucleus, $\xc$, and environment degrees of freedom, $\xenv$. The splitting $(\xtot, \xc)$ allows us to carry out the nonequilibrium statistical mechanics more straightforward as $\xtot$ is a conserved quantity. However, the state of the ambient phase $\xenv$ is not conserved as it exchanges particles, momentum and energy with the nucleus and as such, the nonequilibrium statistical mechanics is slightly more involved with the splitting $(\xenv, \xc)$.

The macroscopic state variables must be computable by exploiting specific microscopic expressions. On this more detailed level of description, the system is characterized by $N$ atoms with mass $m$, position $\rzi$ and momentum $\pzi$ in a box of volume $\Vtot$. All atoms build the microscopic phase space $\bar{\Gamma} = (\bm{r}_1, \bm{p}_1, ..., \bm{r}_N, \bm{p}_N)$. In order to distinguish particles that are part of our nucleus of interest from the remaining ones we introduce the function $\Pi_{\theta(\bar{\Gamma})}$ on $\bar{\Gamma}$. We use the simple position dependent rule $\Pi_{\theta(\Gamma)}(\rzi) = 1$ if particle $i$ is part of the nucleus and $0$ otherwise. Notice that in our notation $\Pi_{\theta(\bar{\Gamma})}(\rzi)$ is seemingly only depending on $\rzi$, but this function of course depends on all position degrees of freedom of $\bar{\Gamma}$. Popular cluster criterions $\Pi_{\theta(\bar{\Gamma})}$ of the kind are the Stillinger-cluster definition\cite{Wedekind:2007adee}, and its extension presented by ten Wolde et al.\ \cite{Wolde:1998,Wedekind:2007adee} (see also Sec.~\ref{Sec_clustering}).  We obtain the subsequent microscopic definition of the macro-variables $\xc$ and $f(\xc)$ characterizing the state of the nucleus:
\begin{equation}
\begin{split}
&\Pi_{\mc} 			= \sum_i \Pi_{\theta(\bar{\Gamma})}(\rzi), \\& 
 \Pi_{\pc}  			= \sum_i \pzi\Pi_{\theta(\bar{\Gamma})}(\rzi), \\&  
 \Pi_{\uc}^i  			= \sum_i \left[\frac{\left(\pzi -m \Pi_{\vc}\right)^2}{2m}  + \phi(\rzi) \right]\Pi_{\theta(\bar{\Gamma})}(\rzi), \\&
\Pi_{f(\xc)}			 = \delta(\Pi_{\xc} - \xc),
\label{microscopic_state_var}
\end{split}
\end{equation}
where $ \phi(\rzi) $ denotes the total interraction potential of particle $i$ due to all other particles. The velocity $\Pi_{\vc}$ of the nucleus is obtained from $\Pi_{\mc} \Pi_{\vc} =\Pi_{\pc}$. By averaging these quantities over phase space with the proper statistical ensemble, we recover the corresponding macrostate variables $\xc$. The variables $\Pi_{\xtot} = (\Pi_{\mtot}, \Pi_{\ptot}, \Pi_{\etot}, \Pi_{\Vtot})$ as the microscopic expressions for $\xtot$ are introduced analogously. Using the shorthand notation $\Pi_{\xc} = (\Pi_{\mc}, \Pi_{\pc}, \Pi_{\uc})$ we also introduced the atomistic expression for the distribution function $f(\xc)$ in Eq.~(\ref{microscopic_state_var}).

\subsection{Nonequilibrium Ensemble} 

Under the asumption that the collection of variables $(\Pi_{\xtot}, \Pi_{f})$ captures all relevant physical processes on the macroscopic time-scale of interest, the nonequilibrium state of the system is characterized by the generalized canonical ensemble\cite{HCO:POT,HCO}: 
\begin{equation}
\begin{split}
\rho_{\xtot, \lambdac} =& \frac{1}{Z(\xtot, \lambdac)} \delta(\Pi_{\xtot} - \xtot) \times \\& \qquad e^{- \int d\xc \lambdac(\xc, \xtot) \Pi_{f(\xc)}   },
\end{split}
\end{equation} 
where $Z$ is the partition function of the mixed microcanonical-canonical ensemble $\rho_{\xtot, \lambdac}$. In particular, the Lagrange multipliers $\lambdac$ must be chosen to guarantee for the proper macroscopic averages of the slow variables, i.e., $f(\xc) = \langle \Pi_{f(\xc)} \rangle_{\rho_{\xtot, \lambdac}}$, where the brackets denote an average over the entire phase space with respect to the ensemble $\rho_{\xtot, \lambdac}$. We also have $\xtot = \langle \Pi_{\xtot} \rangle_{\rho_{\xtot, \lambdac}}$. While the variables $\xtot$ are treated microcanonically since the total system is closed and the quantities $\xtot$ rigorously conserved, the nucleus, as an open system, must be treated canonically through the use of Lagrange multipliers.

\subsection{Entropy} 

We turn our attention to the entropy of the system described by $(\xtot, f(\xc))$. Subsequently, the entropy is shown to generate the irreversible dynamics of $f$. For the sake of clarity, we have summarized all different entropies involved in our analysis in Table~\ref{TABLE_SUMMARY_ENTROPIES}. 

We first derive a useful identity which relies on the observation that the nonequilibrium ensemble can be rewritten as
\begin{equation}
\begin{split}
\rho_{\xtot, \lambdac} = \frac{1}{Z({\xtot, \lambdac})} \delta(\Pi_{\xtot} - \xtot)e^{- \lambdac(\Pi_{\xc}, \xtot)  },
\label{neq_prob}
\end{split}
\end{equation} 
and therefore the distribution $f$ of the nucleus can be cast into the form
\begin{equation}
\begin{split}
f(\xc) &= \langle \Pi_{f(\xc)} \rangle_{\rho(\xtot, \xc)} \\&
	=  \frac{\tilde{Z}({\xtot, \xc})}{Z({\xtot, \lambdac})} \exp\left(- \lambdac(\xc, \xtot)   \right). 
\end{split}
\end{equation} 
Here, $\tilde{Z}$ is the microcanonical counterpart of $Z$,
\begin{equation}
\begin{split}
\tilde{Z}({\xtot, \xc}) &=  \int d\bar{\Gamma} \ \delta(\Pi_{\xtot} - \xtot)\delta(\Pi_{\xc} - \xc),
\label{eq_Ztilde}
\end{split}
\end{equation} 
and we associate it with the entropy ${S}(\xtot, \xc) = \kb \log(\tilde{Z}(\xtot, \xc ))$ of the system in state $\xtot$ with the nucleus of interest having properties $\xc$. We obtain the entropy $S(\xtot, f)$ of the system\cite{HCO:POT,HCO}
\begin{equation}
\begin{split} 
\frac{S(\xtot, f)}{\kb} &= \log(Z({\xtot, \lambdac})) - \int d\xc \lambdac(\xtot, \xtot) f(\xc)  \\&
	     = - \int d\xc \log\left(\frac{e^{-\lambdac(\xtot, \xtot)}}{Z} \right)f(\xc) \\&
	     = - \int d\xc \left[\log\left({f(\xc)}  \right) -\frac{{S}(\xtot, \xc)}{\kb} \right]f(\xc) \\& 
	     =- \int d\xc f(\xc) \log\left(\frac{f(\xc)}{\phi_S(\xtot, \xc)}\right)
 \\&\quad + \frac{\stot(\xtot)}{\kb}, 
\label{entropy_eq}
\end{split}
\end{equation}  
where $ \stot(\xtot) = \kb \log(Z_{\text{tot}})$ is the entropy of the full metastable phase and contains its total partition function
\begin{equation}
\begin{split}
Z_{\text{tot}} &=  \int d\bar{\Gamma} \ \delta(\Pi_{\xtot} - \xtot) \\&
			   = \int_{\xtot} d\bar{\Gamma}. 
\label{eq_Ztot}
\end{split}
\end{equation} 
The contribution $\stot(\xtot)$ will turn out to be irrelevant in generating the final evolution equations. In the last line of Eq.~(\ref{eq_Ztot}) the integration is over the full phase space compatible with the system state $\xtot$. The expression for the entropy $S(\xtot, f)$ contains the entropic potential
\begin{equation}
\begin{split} 
\phi_S(\xtot, \xc) &= \len \delta(\Pi_{\xc} - \xc) \ren_{\xtot} \\&
 \equiv \frac{\int_{\xtot} d\bar{\Gamma} \delta(\Pi_{\xc} - \xc)}{\int_{\xtot} d\bar{\Gamma}}.
\label{entr_pot}
\end{split}
\end{equation}  
The entropic potential $\phi_S$ is related to the probability of occurence of a nucleus with properties $\xc$ in a system described by $\xtot$. Upon studying the dynamics of the distribution $f$ it will turn out that the entropic potential $\phi_S$ represents the nucleation landscape in which the diffusion process describing nucleation takes place, i.e., $V_S(\xtot,\xc) = - \log(\phi_S(\xtot, \xc))$. In the truncated case $\xc^{\text{trc}} = (\mc)$ when the full system is isothermal at temperature $T$ and Brownian motion of the nucleus is irrelevant, the Gibbs-free energy of formation of a nucleus has been recognized\cite{Kusaka:1998b, Oh:1999, Oh:2000,Chen:2001,Yoo:2001,Gonzalez:2014,Wolde:1998} to be $\kb T$ times the logarithm of the probability that a nucleus with size $\mc$ appears in the metastable gas. Our expression $V_S(\xtot,\xc) = - \log(\phi_S(\xtot, \xc))$ hence represents a plausible extension to more nuclei degrees of freedom. In the general setup where the nucleus energy fluctuates, the Gibbs-free energy of formation can no longer be regarded as the proper potential\cite{Schweizer:2014}, but now, $V_S$ or respectively ${S}(\xtot, \xc)$ in Eq.~(\ref{entropy_eq}) as the entropy of formation takes this role. For a collection of the various forms of the nucleation landscape used throughout our work, see Table~\ref{TABLE_SUMMARY_ENTRLANDSCALE}.
\renewcommand{\arraystretch}{2} 
\begin{table} 
\caption{Summary of relevant entropies.}
\begin{adjustwidth}{0cm}{0cm}
\begin{tabular*}{0.48\textwidth}{ p{2.1cm} p{6.3cm} c} 
Expression 			& 
Comments 	  		& 	\\  
\cline{1-2} 			
$S(\xtot, f(\xc)) $				&	 	
Total system entropy when state $\xtot$ and nucleus distribution $f(\xc)$ is known, see Eq.~(\ref{entropy_eq}). Generates dynamics of $f(\xc)$ through Eq.~(\ref{COMPL_GEN_EQ}). 		\\   
$S_{\text{tot}}(\xtot) $				&	 	
Entropy of full metastable phase, including the nucleus with arbitrary state. \\
$S(\xtot, \xc)      $				&	 	
Entropy of system in state $\xtot$ containing nucleus in state $\xc$. 						\\   
$S(\xtot, \mc, \uc)      $			&	
Entropy of system in state $\xtot$ containing nucleus in state $(\mc, \uc)$. Here, the momentum of the nucleus is unknown. See Eq.~(\ref{AMP_ENTR_SPLIT}). 						\\ 
 $S(\xenv, \mc, \uc) $					&
Entropy of system, where the system is split into a nucleus in state $(\mc, \uc)$ and its ambient phase in state $\xenv$. In this set of variables, the entropy can be decomposed: $S(\xenv, \mc, \uc)= \senv(\xenv) + S_c(\mc, \uc) $.						\\   
$\senv(\xenv) $					&
Entropy of nucleus ambient phase to define ambient temperature, pressure and chemical potential, see aSec.~\ref{SEC_ST_VAR}.		\\   
$S_c(\mc, \uc) $					&		   
Entropy of nucleus with state $\xc$ to define nucleus temperature, see also Sec.~(\ref{SEC_tdwlimig}).						\\      
$S_{c,\text{K}}(\mc, T) $			&		 	 
MFKT entropy of nucleus of size $\mc$ in ambient phase at temperature $T$.						\\     
$S_{c,\text{K}}(\mc, \uc) $			&		 	
Extended MFKT entropy of nucleus in state $(\mc, \uc)$.  						\\     
\label{TABLE_SUMMARY_ENTROPIES}
\end{tabular*} 
\end{adjustwidth}
\end{table}%

The entropy of formation ${S}(\xtot, \xc)$ is recognized to depend on the nucleus momentum $\pc$. To highlight this feature we show in App.~\ref{sec_entr_splitt} that in the limit of a sufficiently large nucleus ambient phase,
\begin{equation}
\begin{split} 
 S(\xtot, \xc) =& S(\xtot,\mc, \uc) - \frac{3}{2} \kb\mc \log(\mc) \\&-m \mc\frac{\left(\vc(\mc,\pc) - \bm{v}_{\text{tot}}(\mtot, \ptot) \right)^2}{2 T}, 
\label{AMP_ENTR_SPLIT}
\end{split}
\end{equation} 
where $\bm{v}_{\text{tot}} = \ptot/(m\mtot)$ denotes the overall velocity of the system, $\vc = \pc/(m\mc)$ the nucleus velocity and $S(\xtot,\mc, \uc)$ represents the entropy of formation if only the variables $(\mc,\uc)$ characterize the nucleus. 
 
\subsection{Coarse Grained Evolution Equation}

Coarse-grained evolution equations can be split into a reversible part with the system energy as driving force and a part of irreversible origin involving the entropy as driving force\cite{Schweizer:2014,HCO}. For our system the reversible time-evolution is absent since the only involved dynamics is the exchange of mass, momentum and energy of the nuclei with the ambient phase and these processes are known to be purely irreversible in homogeneous media. The fundamental dynamics is hence encompassed in the irreversible contribution to the evolution equations. In App.~\ref{sec_evolution_eqn_derivation} it is then shown that the evolution equations are recovered as:
\begin{equation}
\begin{split} 
& \frac{\partial f}{\partial t} =  \frac{\partial}{\partial \xc} \cdot \bm{D}(\xtot, \xc) \cdot \left( \frac{\partial f}{\partial \xc} + f \frac{\partial V_S(\xtot, \xc)}{\partial \xc} \right), \\&
   \frac{\partial \xtot}{\partial t} = 0,
\label{f_simple_evolution_eq}
\end{split}
\end{equation}  
where the nucleation landscape $V_S$ drives the dynamics and the elements of the diffusion tensor are
\begin{equation}
\begin{split} 
\bm{D}(\xtot, \xc) = \frac{1}{2 \kb \tau_{\text{GK}}} \len \Delta_{\tau_{\text{GK}}} \Pi_{\xc}\Delta_{\tau_{\text{GK}}} \Pi_{\xc} \ren_{(\xtot, \xc)},
\label{GK_eq}
\end{split}
\end{equation} 
The average $\langle ...\rangle_{(\xtot, \xc)}$ is over all configurations compatible with the state $\xtot$ of the total system and the initial state $\xc$ of the nucleus. Here, $ \Delta_{ \tau_{\text{GK}}} \Pi_{\xc} = \Pi_{\xc}( \tau_{\text{GK}}) -  \Pi_{\xc}(0)$, where $ \tau_{\text{GK}}$ is a time-scale on which the microscopic correlations drop off and the right hand-side in Eq.~(\ref{GK_eq}) reaches a plateau value. For the nucleation of liquid drops in a metastable gas, we recognize $\tau_{\text{GK}}$ as the intermediate time-scale between the collision time of gas particles with the nucleus and the time-scale on which changes in the nucleus state are recognizable on the coarse-grained level of description. We obtain therefore the tensorial generalization (\ref{GK_eq}) of the expression for the diffusion coefficient $(2 \tau_{\text{GK}})^{-1}  \len \Delta_{ \tau_{\text{GK}}} \Pi_{m_c}\Delta_{ \tau_{\text{GK}}} \Pi_{m_c} \ren_{(\xtot, m_c)}$ in Eq.~(\ref{Zeldovic_eq}) already used in a variety of papers \cite{Lundrigan:2009,Auer:2004} for the truncated state space $\xc^{\text{trc}} = (\mc)$. The evolution equation (\ref{f_simple_evolution_eq}) has been discussed in detail by Schweizer and Sagis\cite{Schweizer:2014}, but here we constructed microscopic expressions for its building blocks, $\bm{D}$ and $V_S$.

\subsection{Coarse Grained Evolution Equation in ``Standard" Variables}
\label{SEC_ST_VAR}

In the literature nucleation is often described by decomposing the system into a nucleating phase and its ambient phase\cite{Schweizer:2014,Reguera:2003tt, Reguera:2003ttt}. We now aim to reformulate our theory in terms of the state variables $(\xenv, f)$, where the environment state $\xenv = (\menv, \penv, \uenv,  \Venv)$ is composed of the total number of particles $\menv$, the internal energy $\uenv$, the momentum $\penv$ and the volume $\Venv$ of the nucleus ambient phase. Introducing the average $\langle A \rangle_{f} =  \int d\xc \ f(\xc) A $ of a coarse-grained variable $A(\xc, \xtot)$ with respect to the distribution function $f$, we can relate $\xenv$ to $\xtot$, i.e., 
\begin{equation}
\begin{split} 
& 	\Venv = \Vtot - \langle \Vc(\xc) \rangle_{f}   \\&
	\penv   =  \ptot - \langle \pc \rangle_{f}   \\&
	\uenv   = {\etot - \len \uc + \frac{\pc^2}{2m \mc} \ren_f} - \frac{\penv^2}{2m \menv}, \\&
	\menv ={\mtot - \len \mc \ren_f}.
\end{split}
\end{equation} 
The volume of the ambient phase $\Venv$ is now a time-dependent variable since the volume of the nucleus $V_c(\xc)$ continuously changes during the nucleation process. In terms of the new variables the entropy $\tstot(\xenv, \mc, \uc) $ with eliminated momentum degree of freedom, see also Eq.~(\ref{AMP_ENTR_SPLIT}), can naturally be used to define the temperature $T$ of the ambient phase as the derivative of $\tstot$ with respect to $\uenv$ at fixed nucleus state, i.e., ${1}/{T} = {  \partial \tstot(\xenv, \mc, \uc)}/{  \partial \uenv}$. Analogously, the chemical potential $\mu$ and the pressure $P$ are given by ${\mu}/{T} =-{  \partial \tstot(\xenv, \mc, \uc)}{  \partial \menv}$ and ${p}/{T} =  {  \partial \tstot(\xenv, \mc, \uc)}/{  \partial \Venv}$. If we now assume that in the variables $(\xenv, \xc)$ the entropy splits into its environment and nucleus part, i.e.,  $S(\xenv,\mc, \uc)  = S_{\text{env}}(\xenv) + S_c(\mc, \uc)$, where $S_{\text{env}}(\xenv)$ is a property of the ambient phase only and similarily the nucleus entropy $S_c(\mc, \uc)$ depends solely on the nucleus state, we find upon transforming the evolution equation (\ref{f_simple_evolution_eq}) to the new variables,
\begin{equation}
\begin{split}
\frac{\partial f}{\partial t} &=
 \frac{\partial}{\partial \xc} \cdot \bm{D}(\xenv, \xc) \cdot  
\left(
	\frac{\partial f}{\partial \xc}  
	+f \frac{\partial V_S(\xenv,\xc)  }{\partial \xc} 
\right). 
\label{simple_FPeq2}
\end{split}
\end{equation}
Nucleation takes now place in the effective nucleation barrier
\begin{equation}
\begin{split}
 {\kb}  &V_S(\xenv,\xc) =-{S_c(\mc, \uc)}
		+ \Vc \frac{p}{ T} - \mc \frac{\mu}{ T}  + \frac{\uc}{T} \\& + \mc \frac{\left(\vc(\mc,\pc) - \venv(\menv, \penv) \right)^2}{2T} +\frac{3}{2}{\kb}  \log(\mc),
\label{Veff_eqn}
\end{split}
\end{equation}
where $\venv = \penv/(m \menv)$ is the velocity of the ambient phase. The expression (\ref{Veff_eqn}) was already recovered in the nonequilibrium thermodynamics description of the nucleation process and has been discussed in full detail \cite{Schweizer:2014}. Our statistical mechanics approach however produces the additional logarithmic contribution in $\mc$ and the important observation is, that the introduction of the momentum $\pc$ into the nucleus state space must be accompanied by this logarithmic term in the nucleation barrier.

\subsection{Entropic Potential in Thermodynamic Limit}
\label{SEC_tdwlimig}

Sampling the entropic potential $\phi_S(\xtot, \xc)$ requires to calculate the probability of occurence of a nucleus with properties $\xc$ in the microcanonical ensemble defined by the state $\xtot$. In the prominent case of a sufficiently large system ensemble averages become identical and instead of averaging over the microcanonical ensemble defined by $\xtot$ we can sample over the isobaric-isothermal ensemble with fixed number of particles $N$, fixed pressure $P$ and temperature $T$. By Galilean invariance we can set $\ptot = 0$ without loss of generality. We define the corresponding system state by $\txtot = (N, P, T, \ptot = 0)$. Averages of a phase-space variable $\Pi_{A}$ with respect to the isobaric-isothermal ensemble are given by
\begin{equation}
\begin{split}
\langle \Pi_{A} \rangle_{\txtot} = \frac{\int d\bar{\Gamma} \ \Pi_{A} e^{-\frac{1}{\kb T} \left(\Pi_{\etot} - P \Pi_{\Vtot} \right)}}{\int d\bar{\Gamma} \ e^{-\frac{1}{\kb T} \left(\Pi_{\etot} - P \Pi_{\Vtot} \right)}}. 
\label{NPT_ens_av_eq}
\end{split}
\end{equation}
For the entropic potential (\ref{entr_pot}) we hence obtain the alternative expression
\begin{equation}
\begin{split}
\phi_S(\txtot, \xc) &= \len \delta(\Pi_{\xc} - \xc) \ren_{\txtot},
\end{split}
\end{equation}
valid when the nucleus ambient phase is sufficiently large.  By the same argument other ensembles can be used to sample $\phi_S$, such as the canonical or grandcanonical. Only when the system size is small, the rigorous exact version involving microcanonical averaging must be used.
\renewcommand{\arraystretch}{2} 
\begin{table} 
\caption{Summary of relevant nucleation landscapes.}
\begin{adjustwidth}{0cm}{0cm}
\begin{tabular*}{0.48\textwidth}{ p{2.1cm} p{6.3cm} c} 
Expression 			& 
Comments 	  		& 	\\  
\cline{1-2} 			
$V_S(\xtot, \xc) $				&	 	
Exact landscape valid for any system size, $-\log \len \delta(\Pi_{\xc} - \xc) \ren_{\xtot} $.		\\   
$V_S(\xenv, \xc) $				&	 	
Exact landscape when the system entropy can be split into the nucleus entropy and ambient phase entropy, see Eq.~(\ref{simple_FPeq2}). Used to extend MFKT to richer nucleus state space.	\\   
$V_S(\txtot, \xc) $				&	 	
Approximate Landscape valid when nucleus ambient phase is large, see Eq.~(\ref{eq_final_barr}). Used to study nucleation when nucleus environment is isothermal and isobaric.	\\   
$V_S(\txtot, \mc) $			&	
Approximate truncated landscape valid when nucleus ambient phase is large. Used to study nucleation when nucleus environment is isothermal and isobaric.						\\ 
 $V_S(\txtot, \bm{x}_{c}^{\text{red}}) $					&
Approximate landscape when nucleus ambient phase is large and the nucleus described by its size and kinetic energy. See Eq.~(\ref{EQ_V_RED}). Leads to temperature fluctuation law of McGraw and Laviolette.						\\     
\label{TABLE_SUMMARY_ENTRLANDSCALE}
\end{tabular*} 
\end{adjustwidth}
\end{table}%
The momentum and position degrees of freedom can be separated, so that the entropic potential can be sampled by MC simulations on the position of the particles. To this extent we integrate out the momentum degrees of freedom and rewrite the entropic potential in a more suitable form:
\begin{equation}
\begin{split}
 \phi_S(\txtot,\xc) &= \int de_c' \len \delta(\Pi_{\mc} - \mc)  \delta(\Pi_{\pc} - \pc)\right. \\& \left. \times \delta\left(\Pi_{u_c^{\text{pot}}} + e_c' -  \uc  \right)\right. \\& \left. \times 
\delta\left[ \left(\Pi_{\ekinct} - \frac{\pc^2}{2m\mc} \right)- e_c'\right]
\ren_{\txtot} \\&
\propto  \int_{0}^{\infty}de_c'F\left(n_c,e_c' , T \right)   \phi_S\left(\txtot, n_c,u_c  - e_c' \right),
 \label{NUM_entr_pot}
\end{split}
\end{equation}
where
\begin{equation}
\begin{split}
F =& \frac{1}{ n^{3/2} \Gamma\left(C_V + 1 \right)}  \left(\frac{e_c' }{\kb T}\right)^{C_V}  e^{-\frac{1}{\kb T} \left( e_c' + \frac{\pc^2}{2mn_c} \right)} 
\end{split}
\end{equation}
and $\phi_S\left(\txtot, n_c,u_c  - e_c' \right)$ denotes the probability that in the isobaric-isothermal ensemble the nucleus of interest has size $n_c$ and potential energy $u_c^{\text{pot}} = u_c  - e_c' $. Here, $e_c'$ can be thought of an integration over all possible internal kinetic energy contributions. We also introduced the atomistic expression $\Pi_{u_c^{\text{pot}}} $ of the potential energy of the nucleus as the sum of the potential energies of all atoms of the nucleus, analogous to Eq.~(\ref{microscopic_state_var}), and the total kinetic energy $\Pi_{\ekinct} $ of the nucleus as the sum of the kinetic energy of all its contained atoms. Furthermore, $C_V = \frac{3}{2}\mc-\frac{5}{2}$ is the heat-capacity of an ideal gas. The probability $\phi_S\left(\txtot, n_c,u_c^{\text{pot}} \right)$ can conveniently be sampled by means of MC simulations since it does not depend on the momentum degrees of freedom.   

\begin{figure}[!htb]
\centering
\includegraphics[width=8.5cm]{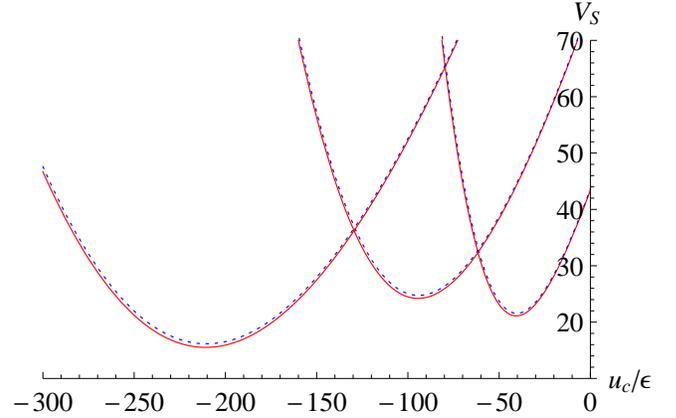}
\caption{Nucleation landscape $V_S(\txtot, \xc)$ of a liquid drop in a metastable argon gas for a nucleus with $\mc =20,40$ and $80$ particles and $\pc = 0$ as a function of its internal energy $\uc$. The red line shows the approximate landscape from Eq.~(\ref{eq_final_barr}) and the dotted blue line the exact landscape with integration of Eq.~(\ref{eq_exactphi}). Thermophysical data is taken from the sixth row in Table~\ref{TABLEPARAM}.}
\label{Fig_approx_comp}
\end{figure}

To gain further insight, we approximate the probability $\phi_S(\txtot, \mc,u_c^{\text{pot}} )$ by the saddle point approximation obtained for each nucleus size $n_c$ in the potential minimum $u_c^{\text{pot},0} (\mc) = \text{min}_{u_c^{\text{pot}}} \phi_S(\txtot, \mc,u_c^{\text{pot}} )$. In this minimum we can evaluate the second derivative of  $\phi_S(\txtot, \mc,u_c^{\text{pot}} )$ with respect to $u_c^{\text{pot}} $, which we denote by $H(n_c)$ and find:
\begin{equation}
\begin{split}
\phi_S(\txtot, n_c,u_c^{\text{pot}} )\propto & \sqrt{H(\mc)}\phi_S(\txtot,\mc)  \\& \times  e^{- H(\mc)   \frac{\left(u_c^{\text{pot}}- u_c^{\text{pot},0}(\mc) \right)^2}{2}  }.
\label{eqn_phi_approx}
\end{split}
\end{equation} 
Here the factor of $ \sqrt{H(\mc)}$ is arising for normalization reasons. The entropic potential $\phi_S(\txtot, \mc)$ of the truncated theory is recovered from
\begin{equation}
\begin{split}
\phi_S(\txtot, \mc) &= \int du_c^{\text{pot}} \ \phi_S(\txtot, \mc,u_c^{\text{pot}}) \\&
	  = \int du_c^{\text{pot}} \len \delta(\Pi_{\mc} - \mc) \delta (\Pi_{u_c^{\text{pot}}} - u_c^{\text{pot}}) \ren \\&
        = \len \delta(\Pi_{\mc} - \mc) \ren
\end{split}
\end{equation} 
and with this we can define the nucleation landscape $V_S(\txtot, \mc) = -\log(\phi_S(\txtot, \mc))$ of the truncated theory. With the approximation (\ref{eqn_phi_approx}) we can solve the integral in Eq.~(\ref{NUM_entr_pot}) analytically. A more elegant way is to rewrite $\phi_S(\txtot, \mc)$ in the form 
\begin{equation}
\begin{split}
 &\phi_S(\txtot,\xc) \propto   \frac{e^{-\frac{\pc^2}{2mn_c \kb T}}}{ \mc^{3/2} \Gamma\left(C_V + 1 \right)} 
\sqrt{H(\mc)}\phi_S(\txtot,\mc) \\& \quad
\times \int_{0}^{\infty}de_c' e^{-H\frac{\left(\uc -e_c'-V_c^{0}  \right)^2}{2} - \frac{e_c'}{\kb T}+ C_V \log\left(\frac{e_c'}{\kb T}\right)}.
\label{eq_exactphi}
\end{split}
\end{equation}
We approximate the exponent in the integral to second order around the saddle point $e_c^{s}$ and we are then able to calculate the resulting Gaussian integral. The lower integration boundary can be shift to $-\infty$ since the main contribution of the integral comes from values $e_c^{'} >0$ so that the nucleation landscape $V_S(\txtot,\xc) = - \log\left(\phi_S(\txtot,\xc) \right)$ is given by
\begin{equation}
\begin{split}
 &{V_S(\txtot,\xc)} =V_S(\txtot, \mc) + \frac{\pc^2}{2m\mc \kb T}  \\&\qquad
+\frac{3}{2}\log(\mc) + \log(\Gamma(C_V + 1)) + H \frac{\left(\uc - e_c^{s} - V_0 \right)^2}{2} \\&\qquad
+\frac{e_c^s}{\kb T} - C_V \log\left(\frac{e_c^s}{\kb T}\right) + \frac{1}{2}\log\left(1+\frac{C_V}{H \left(e_c^s\right)^2}\right) \\&\qquad
+ \text{const}(T)
\label{expanded_pot_eq}
\end{split}
\end{equation}
that turns out to compare excellently with the true solution (\ref{eq_exactphi}) of the integral for nucleus sizes down to $n_c = 5$, see also Fig~\ref{Fig_approx_comp}. 
To gain further insight, we can introduce a thermodynamic definition of the nucleus temperature $T_c$. We recall that $V_S(\xtot,\xc) = - \log(\phi_S(\txtot(\xtot),\xc))$ and upon representing the nucleation landscape $V_S$ in terms of the state of the ambient phase $\xenv = (\menv, \penv (=0), \uenv, \Venv)$, see Sec.~\ref{SEC_ST_VAR}, we obtain 
\begin{equation}
\begin{split}
\frac{\partial V_S(\xtot, \xc)}{\partial \xc} &= \frac{\partial V_S( \xe, \xc)}{\partial \xc} + \frac{\partial \xe }{\partial \xc} \frac{\partial V_S(\xe,\xc)}{\partial \xe} 
					 \\&    = \frac{\partial V_S(\xe,\xc)}{\partial \xc} - \frac{\partial V_S(\xe, \xc)}{\partial \xe}.  
\label{id1}
\end{split}
\end{equation}  
Notice, that the variable set $(\xc, \xtot)$ is more convenient to evaluate the statistical mechanics, but the physical interpretation of $V_S$ is more appealing in terms of $(\xc, \xe)$. When changing the nucleation landscape $V_S( \xe, \xc)$ with respect to the internal energy $\uenv  $ of the ambient phase, while fixing the nucleus properties $\xc$, this change must be equal to the temperature of the environment, such that ${\partial V_S(\xe,\xc)}/{\partial U_{\text{e}}} = -{1}/{\kb T}$, where the minus sign is a result of our sign convention entering the definition of $V_S$ through $V_S = -  \log(\phi_S)$, see also the form of $V_S$ presented in Eq.~(\ref{Veff_eqn}). Similarily, the nucleus temperature can be assigned as ${\partial V_S(\xe,\xc)}/{\partial \uc} = -{1}/{\kb T_c}$. The change of variables law (\ref{id1}) implies 
\begin{equation}
\begin{split}
\frac{\partial V_S( \xtot, \xc)}{\partial u_c} &= \frac{1}{\kb T} - \frac{1}{\kb T_c}  
\label{UC_eqn}
\end{split}
\end{equation}
and employing now the specific expression (\ref{expanded_pot_eq}) for the purpose of evaluating the left hand-side of Eq.~(\ref{UC_eqn}) we find the caloric equation of state,
\begin{equation}
\begin{split}
 \uc =  C_V \kb T_c + V_c^0 + \frac{1}{\kb H} \left(\frac{1}{T} - \frac{1}{T_c}\right),
\label{uc_const_eqn}
\end{split}
\end{equation} 
and with this, the saddle point can be represented as $e_c^s = C_V T_c$. To obtain Eq.~(\ref{uc_const_eqn}) we ignored the second last term in Eq.~({\ref{expanded_pot_eq}) since compared to the other terms it is of subleading order in $\mc$. The internal energy of a nucleus is hence given by the kinetic contribution $C_V T_c$ and a potential energy contribution involving the most probable potential energy $V_c^0$ at a given nucleus size $\mc$, and a correction for nonisothermal effects. When $T =T_c$, the nucleus is in the potential minimum of the landscape $\phi_S(\txtot,\xc)$ at fixed $\mc$ and $\pc$. Plugging Eq.~(\ref{uc_const_eqn}) and $e_c^s = C_V T_c$ back into the expression (\ref{expanded_pot_eq}) for the entropic potential and expanding the logarithm of the $\Gamma$-function with Stirling's approximation, we obtain the final form for the nucleation landscape
\begin{equation}
\begin{split}
 {V_S(\txtot, \xc)} =&  V_S(\txtot, \mc)  +\frac{\pc^2}{2Tn_c}  \\&
  			-C_v \left[\log\left(\frac{T_c}{T} \right) + \left(1-\frac{T_c}{T} \right) \right] \\& 
+ \frac{1}{2 H} \left(\frac{1}{T}- \frac{1}{T_c}\right)^2+ 2\log(n_c) 
\\&
			 + \mathcal{O}\left(\frac{1}{n_c}\right) + \text{const}(T).
 \label{eq_final_barr}
\end{split}
\end{equation} 
Note, $\phi_S(\txtot, n_c)$ is the usual barrier to nucleation in the absence of Brownian-motion and non-isothermal effects. In particular, $3/2 \log(n_c)$ of the correction are due to the introduction of the momentum $\pc$ into the nucleus state space and $1/2\log(n_c)$ through the Stirling approximation on the Gamma function $\Gamma(C_V + 1)$. We here recover all corrections reported in the literature, including the momentum contribution obtained through entirely thermodynamic arguments by Reguera and Rubi\cite{Reguera:2001q}, and the temperature fluctuation law noticed by McGraw and Laviolette\cite{McGraw:1995}. The additional temperature fluctuation contribution proportional to $1/H$ has not been reported in the literature to our knowledge. In the subsequent studied model system of argon nucleation the contribution proportional to $1/H$ effects the energetic landscape of $V_S(\txtot, \xc)$ dramatically and is hence relevant (see Fig.~\ref{Fig_approx_comp}). We postpone further discussion of the temperature fluctuations to Sec.~\ref{sec_T_fluct}.

\section{Kinetic Energy as a State-Variable}
\label{Sec_kin_e_as_statevar}

In view of the various equivalent expressions for the nucleation landscape $V_S$, see Eq.~(\ref{Veff_eqn}) and Eq.~(\ref{eq_final_barr}) for instance, it is plausible to ask wether the kinetic energy of the nucleus $\ekinc$ can be used as a state variable instead of the nucleus momentum $\pc$, since the latter enters the entropic barrier only in the contribution $\pc^2/2m\mc$ being the ``systematic" kinetic energy of the nucleus. Suppose for the further considerations that $\ptot = 0$ by Galilean invariance. If we use the modified set of nucleus state variables $\xc^{\text{mod}} = (\mc, \ekinc, \uc)$ instead of $\xc = (\mc, \pc, \uc)$ we have to construct the entropic potential in terms of these new variables. The entropic potential $\phi(\xtot,\xc)$ is the probability that we encounter a nucleus with state $\xc$ and therefore we find the clarrifying relationship
\begin{equation}
\begin{split}
 \phi(\xtot,\xc^{\text{mod}} ) = \int d^3\pc \ \delta\left(\ekinc - \frac{\pc^2}{2m\mc}\right) \phi(\xtot,\xc).
\end{split}
\end{equation} 
This can be used to obtain the landscape $V_S(\xtot,\xc^{\text{mod}}) = -\log( \phi(\xtot,\xc^{\text{mod}} )$ which looks exactly like $V_S(\xtot,\xc)$, but with $\pc^2/2m\mc$ replaced by the new independent state variable $\ekinc$. The modified set of state variables have the advantage that the state space is smaller upon catching the same relevant macro-physics. The intuition is that in a homogeneous medium it must be irrelevant in which direction the momentum $\pc$ of the nucleus points and therefore the nucleation process can equally be described by $|\pc|$, $\pc^2$ or the kinetic energy $\ekinc$.

\section{Kinetic Energy Fluctuations in Literature}
\label{sec_T_fluct}

It has been convincingly demonstrated in MD simulations\cite{Wedekind:2007aa,Diemand:2014ads} of nucleation that when only the properties $\xc^{\text{red}} = (\mc, \ekinct)$ of a nucleus are measured, where now $\ekinct$ is the total kinetic energy composed of internal and systematic contributions, the nucleus temperature $T_c$ can be defined by $T_c = \ekinct/C_V$ and its fluctuations obey the law derived by McGraw and Laviolette, $\text{exp}\{C_v \left[\log\left({T_c}/{T} \right) + \left(1-{T_c}/{T} \right) \right]\} $, without the additional contribution involving $H$ we obtained in Eq.~(\ref{eq_final_barr}). We first remark, that the state space $\xc^{\text{red}}$ is a reduced version of $\xc$ and therefore cannot capture entirely the rich physics contained in $\xc$. In $\xc$ we have additional knowledge of the potential energy contributions to the nucleus internal energy and it is then natural, that these contributions give rise to the more complicated caloric equation of state (\ref{uc_const_eqn}) rather than 
$\ekinct = C_V T_c$. Thus, in the reduced framework $\xc^{\text{red}}$ we measure kinetic energy fluctuations only, and define the temperature therein while in the richer theory $\xc$ we measure the internal energy fluctuations composed of kinetic and potential energy fluctuations, and so it is not surprising that the fluctuation law contains additional contributions.

We apply a similar reduction of degrees of freedom as in Sec.~\ref{Sec_kin_e_as_statevar} to show that in the reduced state space $\xc^{\text{red}} $ where only the size and total kinetic energy is measured in a simulation, the temperature indeed obeys the equation of state $\ekinct = C_V T_c$ with the fluctuation law by McGraw and Laviolette. To this extent, notice that the integrand in Eq.~(\ref{NUM_entr_pot}), denoted by $\phi(\txtot, \xc, \ekinct)$ represents the probability that we observe a nucleus with property $\xc$ and total kinetic energy $\ekinct$. Therefore 
\begin{equation}
\begin{split}
 \phi(\txtot,\xc^{\text{red}} ) = \int d^3\pc d\uc \ \phi(\txtot, \xc, \ekinct).
\label{eq_McGraw1}
\end{split}
\end{equation} 
Employing the approximation (\ref{eqn_phi_approx}) the probability $\phi(\txtot, \xc, \ekinct)$ has the form of the right hand side of Eq.~(\ref{eq_exactphi}), but without the integration over $e_c'$. Therefore the integrations in Eq.~(\ref{eq_McGraw1}) can be performed:
\begin{equation}
\begin{split}
 \phi(\txtot,\xc^{\text{red}} ) \propto  \frac{\phi(\txtot, \mc) }{\Gamma(C_V + 1)} e^{-\frac{\ekinct}{\kb T} + C_V \log\left( \frac{\ekinct}{\kb T} \right)}.
\end{split}
\end{equation} 
so that the nucleation landscape $V_S$ of the reduced theory satisfies up to irrelevant constants
\begin{equation}
\begin{split}
V_S(\txtot,\xc^{\text{red}} ) =& V_S(\txtot, \mc)  + \log(\Gamma(C_V + 1)) \\& -\frac{\ekinct}{\kb T} + C_V \log\left( \frac{\ekinct}{\kb T}   \right).
\end{split}
\end{equation} 
The temperature can be defined analogously to the procedure in Sec.~\ref{SEC_tdwlimig} and we obtain $\ekinct = C_V T_c$ and by employing the Stirling approximation finally the suggested fluctuation law by McGraw and Laviolette,
\begin{equation}
\begin{split}
V_S(\txtot,\xc^{\text{red}} ) =&V_S(\txtot, \mc)  +\frac{1}{2}\log(n_c)  \\& -C_v \left[\log\left(\frac{T_c}{T} \right) + \left(1-\frac{T_c}{T} \right) \right].
\label{EQ_V_RED}
\end{split}
\end{equation} 
This shows that the kinetic energy of the nucleus is indeed a valid measure to obtain the nucleus temperature if no information about its potential energy is available.

\section{Application: Extension of MFKT}
 \label{sec_MFKT}

At the example of the MFKT of nucleation\cite{KalikmanovB2013} we illustrate that our work represents a powerful tool to extend existing nucleation theories to richer state spaces. 

In his work\cite{Kalikmanov:2006}, Kalikmanov derived a non-perturbative expression for the partition function $Z_{\text{c,K}}(\mc,T)$ of a nucleus by microscopic considerations. In his treatment the ambient phase acts as a mean-field on the nucleus. His theory is valid for isothermal nucleation at temperature $T$ and in particular he finds
\begin{equation}
\begin{split}
\log(Z_{\text{c,K}}(\mc,T))=&  -\mc \frac{\mu_{\text{sat}}(T)}{T} \\&  -\kb \left[\mc^s(\mc, T) -1\right] \thetam(T) \\&+ \kb\log(\Vtot \rhosat(T)),
\end{split}
\end{equation}
where the chemical potential $\mu_{\text{sat}}(T)$ and particle density $\rhosat(T)$ at saturation are introduced. Here, $\mc^s$ can be interpreted as the number of particles located at the surface of the nucleus and $\thetam$ is a ``microscopic" surface tension. 

The entropy $S_{c,\text{K}}(\mc, T)$ and internal energy $\uc(\mc, T)$ of the nucleus, see also Sec.~\ref{SEC_ST_VAR}, are then given by
\begin{equation}
\begin{split}
& S_{c,\text{K}}(\mc, T) = \kb \log(Z_{\text{c,K}}(\mc,T)) + \kb T \frac{\partial Z_{\text{c,K}}(\mc,T)}{\partial T}, \\&
\uc(\mc, T) = \kb T^2  \frac{\partial Z_{\text{c,K}}(\mc,T)}{\partial T},
\end{split}
\end{equation}
by means of the usual definition of statistical mechanics.

\subsection{Enhancing State Space of MFKT}

We exploit a local equilibrium hypothesis to extend the entropy obtained by Kalikmanov to nonisothermal nucleation. To this extent, notice that the entropy $S_{c,\text{K}}(\mc,T) $ of a nucleus can be viewed as a property of the nucleus only and in this sense does not depend on the state of the environment. In isothermal nucleation the temperature of the nucleus $T_c$ must coincide with the ambient temperature, so that $T = T_c$ and we can equally write $S_{c,\text{K}}= S_{c,\text{K}}(\mc,T_c) $. Now the local equilibrium hypothesis enters and we postulate that a nucleus in nonisothermal nucleation that can be assigned the temperature $T_c$ has the same properties as a nucleus in isothermal nucleation that takes place at temperature $T = T_c$. Since the entropy, and analogously the internal energy of the nucleus are solely a property of the nucleus and do only depend on the nucleus state, we hence find the generalization of the nucleus entropy to describe nonisothermal nucleation,   
\begin{equation}
\begin{split}
& S_{c,\text{K}}(\mc, T_c) = \kb \log(Z_{\text{c,K}}(\mc, T_c)) + \kb  T_c \frac{\partial Z_{\text{c,K}}(\mc, T_c)}{\partial  T_c}, \\&
\uc(\mc,  T_c) = \kb  T_c^2  \frac{\partial Z_{\text{c,K}}(\mc, T_c)}{\partial  T_c},
\end{split}
\end{equation}
where the nucleus temperature $T_c$ is now related to its internal energy and mass by ${1}/{T_c(\mc, \uc)} = {\partial S_{\text{K}}^{\text{ext}}}/{\partial \uc}$. This then implies $S_{c,\text{K}}^{\text{ext}}(\mc, \uc) = S_{c,\text{K}}^{\text{ext}}(\mc, T_c(\mc,\uc))$. 

\subsection{Analytic Solution for Nucleation Rate}

The entropy $S_{\text{K}}^{\text{ext}}(\mc, \uc)$ does not depend on the momentum variable $\pc$, so that we use the equations stated in Sec.~\ref{SEC_ST_VAR} to obtain the dynamics of extended MFKT. Our description then also takes naturally Brownian motion into account. We take the diffusion tensor of the particular form\cite{Schweizer:2014} 
\begin{equation}
\bm{D}(\mc) = 
 \begin{pmatrix}
  \xi(\mc) \mathbf{I}_{3\times 3}        & 0  \\
  0   & \tilde{\bm{D}}         \\ 
 \end{pmatrix},
\label{D_matr}
\end{equation}
and $\tilde{\bm{D}}$ is a $2\times2$ matrix with elements $\tilde{\bm{D}}_{\mc,\mc}$, $\tilde{\bm{D}}_{\mc,\uc}$ and $\tilde{\bm{D}}_{\uc,\uc}$ regulating the mass and heat exchange mechanism between nucleus and ambient phase. The evolution equations (\ref{simple_FPeq2}) of a nucleus in a metastable phase at temperature $T$ and pressure $P$ can be solved in the saddle point approximation\cite{Schweizer:2014,Trinkaus:1983,Barrett:1994} and give the prediction $J_{\text{K, ext}}$ of our extended MFKT:
\begin{equation}
\begin{split}
{J_{\text{K, ext}}}  
=&   \sqrt{-\frac{1}{2\pi}\left.\frac{\partial^2 V_S(\txtot, \xc)}{\partial \mc^2} \right|_{\xc^{*}}} \rho_{g, \txtot}  \\& \times   \frac{1}{ \omega} \frac{|\lambda_{\text{min}}|}{\sqrt{|\text{det} \bm{G} |}}  e^{-\Delta V_S(\txtot, \xc^{*})},,
\label{eq_JMFKTratio}
\end{split}
\end{equation}
where 
\begin{equation}
\begin{split}
\bm{G} = \frac{\tilde{\zeta}^2}{\sigma_{\uc}^2} 
\begin{pmatrix}
  1 - \frac{\omega^2}{\tilde{\zeta}^2} & \frac{1}{\tilde{\zeta}} \\
  \frac{1}{\tilde{\zeta} }& \frac{1}{\tilde{\zeta}^2}
 \end{pmatrix}.
\end{split}
\end{equation}
The coefficients are given by
\begin{equation}
\begin{split}
& \sigma_{\uc}^2 = T^2 \left. \frac{\partial \uc(\mc, T)}{\partial T}\right|_{\mc^{*}}, \\&
 \frac{1}{\sigma_{\mc}^2} = - \thetam(T) \left.\frac{\partial^2 n_s(T,\mc)}{\partial \mc^2}\right|_{\mc^{*}}, \\&
 \omega^2 = \left.\frac{\sigma_{\uc}^2}{\sigma_{\mc}^2}\right|_{\mc^{*}},\\&
 \tilde{\zeta} = -c_v T + T^2 \frac{\partial \log(\rhosat(T))}{\partial T} 
		 + T^2 \left. \frac{\partial^2 n_s(T,n)}{\partial \mc \partial T}\right|_{\mc^{*}}\thetam \\& \qquad 
		 + T^2 \left|\frac{\partial n_s(T,n)}{\partial \mc} \right|_{\mc^{*}}\frac{d \thetam(T)}{dT},
\label{eq_MFKTCOEFF}
\end{split}
\end{equation}
and 
\begin{equation}
\begin{split}
\lambda_{\text{min}} = \frac{1}{2} \left(\text{tr} \bm{R} - \sqrt{(\text{tr} \bm{R})^2 - 4 \text{det} \bm{R}} \right). 
\end{split}
\end{equation}
identified as the smallest eigenvalue of $\bm{R} = \tilde{\bm{D}} \bm{G}$. All quantites in Eq.~(\ref{eq_MFKTCOEFF}) and $\lambda_{\text{min}}$ must be evaluated at the saddle point $\xc^{*} = (\mc^{*}, \pc^{*} = 0, \uc^{*})$ of the nucleation landscape $V_S(\txtot, \xc)$ as given by Eq.~(\ref{Veff_eqn}). For the internal energy $\uc(\mc, T_c)$ this saddle point falls on $\uc^{*} = \uc(\mc^{*}, T)$. Here, $\Delta V_S(\txtot, \xc^{*})$ denotes the height of the nucleation landscape and $\rho_{g} $ is the particle density of the metastable phase in which nuclei are formed. In the saddle point approximation (\ref{eq_JMFKTratio}) the effects of Brownian motion are seemingly irrelevant since the coefficient $\xi$ is not involved. This is however not true since the logarithmic normalization factor $3/2 \log(\mc^{*})$ contained in $\Delta V_S(\txtot, \xc^{*})$ enters the prediction. 

For the prediction of the standard isothermal MFKT nucleation rate $J_\text{K}$ we refer to the book of Kalikmanov\cite{KalikmanovB2013} (Chapter 7.7 Steady State Nucleation Rate).

\subsection{Application to Nucleation of Argon}
\label{sec_Argon_MFKT}

As a specific model system we consider the nucleation of argon droplets in a supersaturated gas in the low-supersaturation regime studied by Diemand et al.\ through brute-force MD simulations\cite{Diemand:2013}. This allows to compare our predictions directly with first hand exact results on the nucleation rate. For the physical setups listed in Table~\ref{TABLEMFKT} we have computed the isothermal MFKT nucleation rate $J_\text{K}$ and the corrected rate $J_{\text{K, ext}}$ and compared it to the MD simulations by Diemand. Since we compared our nucleation rate predictions to MD simulations, we did not take the real thermophysical data of argon, but data obtained through MD simulations\cite{Diemand:2013}, see also Table~\ref{tab1}. 

In Fig.~\ref{Figlocequilibriutemp} we illustrated the predictions of the standard isothermal MFKT and the extended version. The truncated theory seems to deviate systematically from the exact MD rates $J_{\text{MD}}$. Our theory succesfully eliminates this deviation indicating that we chose the proper set of variables to capture the nucleation process on the macroscale. Notice, that we did not include effects of angular momentum and shape fluctuations in our description. As argued by Ang\'{e}lil et al.\ \cite{Diemand:2014ads} the effect of angular momentum seems only to be pronounced for small nuclei consisting of less than $16$ atoms and all our critical sizes were beyond this size, so that angular momentum should not be used as an independent state variable for an autonomous description of the nucleation process in the low supersaturation regime. Ang\'{e}lil et al.\ recognized that the shape of large nuclei can deviate significantly from spherical objects. Using the shape of a nucleus as an independent fluctuating variable might offer additional insights into the relevance of this correction.

Specifically, we evaluated the MFKT nucleation rate by resampling to the following relationships for which precise MD thermophysical data was available, see Table~\ref{TABLETHERMOPHYS}. We shortly summarize the parameters appearing in the MFKT and closely follow the lines of its inventor\cite{Kalikmanov:2006,KalikmanovB2013}. According to Kalikmanov, the microscopic surface tension is given by $\thetam(T_c)  = -\log(-B_2(T_c) \psat(T_c)/(\kb T_c)  )$, where $B_2$ is the second virial coefficient of the gas. Introducing the macroscopic surface tension $\gamma_S(T_c)$, we can define the ratio $\tilde{\omega}(T_c) = \gamma_S(T_c)/(\kb T_c\thetam(T_c))$. We further define $\tilde{\zeta} = n_c^{-1/3}$ and the coefficient $\tilde{\lambda} = \sqrt{N_1/\tilde{\omega} - 3/4} - 3/2$, where $N_1$ is the number of neighbours of particles in the nucleus. The number of neighbours in the nucleus is evaluated from the argon-liquid properties and can be approximated by $N_1(T_c) = 5.5116 \eta(T_c)^2 + 6.1383 \eta(T_c) +1.275$, where $\eta(T_c) = \frac{\pi}{6} \rho_l(T_c) d_{\text{hs}}^3$. Here, $ d_{\text{hs}}$ is given by the Barker-Henderson relation with the Weeks-Chandler-Anderson decomposition of the intermolecular potential, and represents the effective hard sphere diameter in the theory of liquids, $ d_{\text{hs}}(T_c) = {a_1 T + b_1}/({a_2 T_c + a_3}) \sigma$ with the Lennard-Jones parameters $\sigma =3.822\times10^{-10}\,\text{m}$ and $\epsilon = 1.654\times10^{-21}\,\text{J}$. The coefficients are given by $a_1 = 0.56165 \frac{\kb}{\epsilon}$, $a_2 = 0.60899 \frac{\kb}{\epsilon}$, $a_3 = 0.92868 \frac{\kb}{\epsilon}$ and $b_1 = 0.9718 \frac{\kb}{\epsilon}$. The mass of argon particles is $m = 6.690\times10^{-26}\,\text{kg}$. The average number of particles $\mc^s(n_c,T_c) = \tilde{\alpha}(n_c, T_c)n_c$ located at the surface of the nucleus is obtained by solving a cubic equation for the introduced coefficient $\alpha(n_c, T_c)$. In particular, $ \tilde{\alpha} =    3\omega \tilde{\zeta} (1-\alpha)^{2/3} + 3 \tilde{\omega} \tilde{\lambda }\tilde{\zeta}^2 (1-\alpha)^{1/3} + \tilde{\zeta}^3 \tilde{\omega} \tilde{\lambda}^2$.
\begin{figure}[!htb]
\centering
\includegraphics[width=8.5cm]{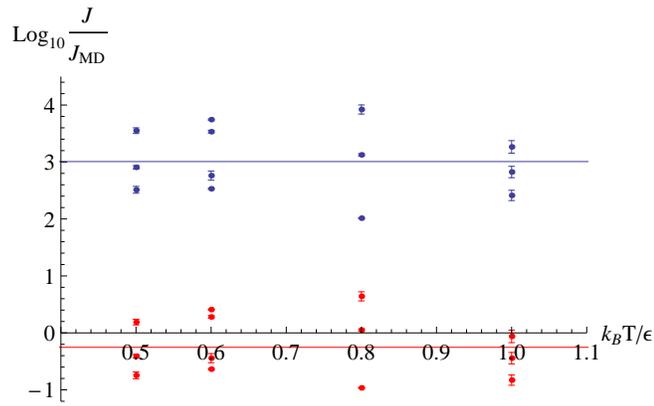}
\caption{ Comparison of argon nucleation rates $J$ obtained from recent large-scale molecular dynamics simulations\cite{Diemand:2013}, $J_{\text{MD}}$, with the original mean-field kinetic theory\cite{Kalikmanov:2006} (blue) and our extension (red). At each temperature $T$ the rate comparison for various supersaturation ratios $S$ is presented, see also Table~\ref{TABLEMFKT}. The solid lines are a guide to the eyes. }
\label{Figlocequilibriutemp}
\end{figure} 
For the kinetic coefficient $\tilde{\bm{D}}_{\mc,\mc}$ we used the standard formula obtained from kinetic theory\cite{Barrett:1994,Barrett:2008}
\begin{equation}
\begin{split} 
\tilde{\bm{D}}_{\mc,\mc}= \frac{\psat}{\sqrt{2\pi m \kb T}} A(\mc),
\label{kinetic_theory_Dnn}
\end{split}
\end{equation}
where $A(\mc) = 4\pi r_c^2$ is the surface area of the nucleus and $r_c =(3\rho_l\mc/4\pi)^{1/3}$ its radius. The remaining coefficients are given by predictions of Barret \cite{Barrett:1994, Schweizer:2014}, i.e., $\tilde{\bm{D}}_{\mc,\uc} = 2\kb T\tilde{\bm{D}}_{\mc,\mc}$ and $\tilde{\bm{D}}_{\uc,\uc} = 6(\kb T)^2\tilde{\bm{D}}_{\mc,\mc}$ and are a result of kinetic considerations of the microscopic mass and energy exchange process between ambient phase and nucleus.
\renewcommand{\arraystretch}{1.5} 
\begin{table} 
\caption{\label{tab1} Thermophysical data of argon\cite{Diemand:2013}.  }
\begin{adjustwidth}{1.25cm}{-1cm}
\begin{tabular*}{0.48\textwidth}{cccccc} 
$\kb T/\epsilon$ 			& 	
$\psat \sigma^3/\epsilon$ 		&
$\gamma_S \sigma^2/\epsilon$		&
$\rho_l \sigma^3/m$  			&	
$B_2/\sigma^3$  			&	\\  
\cline{1-5}
$1.0$	&$2.55 \times 10^{-2}$	&$0.453$& $0.696 $&$-5.26 $		\\   
$0.8$	&$4.53 \times 10^{-3}$	&$0.863$& $0.797 $&$-7.75 $		\\   
$0.6$	&$2.54 \times 10^{-4}$	&$0.882$& $0.882 $&$-12.9$	 	\\   
$0.5$	&$2.54 \times 10^{-5}$	&$0.921$& $0.921 $&$-18.15$		\\      
\label{TABLETHERMOPHYS}
\end{tabular*} 
\end{adjustwidth}
\end{table}%
\renewcommand{\arraystretch}{1.5} 
\begin{table} 
\caption{ Nucleation rate predictions of argon at various temperature $T$ and supersaturations $S$. The initial density of the metastable gas is $\rho_g$. Shown are the predictions $J_{\text{MD}}$ of the brute-force MD simulations\cite{Diemand:2013} by Diemand et al.\, the predictions of the original MFKT\cite{Kalikmanov:2006}, $J_{\text{K}}$, and the predictions, $J_{\text{K,ext}}$ of our extension of MFKT. }
\begin{tabular*}{0.48\textwidth}{ccccccc} 
$\kb T/\epsilon$ 					& 	
$S$ 							&
$\rho_{g} \sigma^3$				&
$\log_{10}\left(\frac{J_{\text{MD}}}{\sigma^{3}\tau}\right)$			&
$\log_{10}\left(\frac{J_{\text{K}}}{\sigma^{3}\tau}\right)$  			&
$\log_{10}\left(\frac{J_{\text{K,ext}}}{\sigma^{3}\tau}\right)$ 			&		\\  
\cline{1-6}
$0.5$	&$72.8$	&$0.008$& $-11.36\pm 0.03$&$-9.15  $ & $-12.14 $		\\   
$0.5$	&$59.2$	&$0.007$& $-13.01\pm 0.02$&$-10.41 $ & $-13.44 $		\\   
$0.5$	&$48.7$	&$0.005$& $-15.07\pm 0.04$&$-11.83 $ & $-14.92 $	 	\\   
$0.6$	&$16.9$	&$0.008$& $-11.75\pm 0.01$&$-9.53  $ & $-12.42 $		\\   
$0.6$	&$15.6$	&$0.007$& $-12.64\pm 0.02$&$-10.17 $ & $-13.10 $		\\   
$0.6$	&$14.0$	&$0.006$& $-14.39\pm 0.05$&$-11.16 $ & $-14.15 $		\\   
$0.6$	&$12.0$	&$0.005$& $-16.32\pm 0.73$&$-12.88 $ & $-15.93 $		\\   
$0.8$	&$4.02$	&$0.030$& $- 9.08\pm 0.01$&$-7.36  $ & $-10.08 $		\\   
$0.8$	&$3.55$	&$0.025$& $-11.70\pm 0.01$&$-8.87  $ & $-11.70 $		\\   
$1.0$	&$1.66$	&$0.062$& $-10.94\pm 0.01$&$-8.83  $ & $-11.79 $		\\   
$1.0$	&$1.63$	&$0.060$& $-11.90\pm 0.07$&$-9.38  $ & $-12.36 $		\\   
$1.0$	&$1.60$	&$0.058$& $-12.92\pm 0.07$&$-9.96  $ & $-13.00 $  		\\    
\label{TABLEMFKT}
\end{tabular*} 
\end{table}%

\section{Application: Nucleation in Lennard-Jones Gas}
\label{sec_LJNucleation} 

We study the nucleation of liquid drops in a metastable Lennard-Jones gas in the low supersaturation regime considered by the latest large-scale MD simulations\cite{Diemand:2013} by Diemand et al.\. In order to compare our nucleation rate predictions obtained from our coarse-graining procedure with the MD simulations, we employ the same Lennard-Jones parameters as Diemand et al.\ and consider here nucleation in the limit of a large ambient phase at temperature $T$ and pressure $P$. We aim to obtain the static building block $\phi_S(\xc, \txtot)$ and the dynamic information $\bm{D}_{\xc,\xc}$ from atomistic simulations. This then determines the coarse-grained model. We apply the following scheme to obtain predictions of the nucleation rate:
\begin{enumerate}
  \item Sample configurations $C$ distributed according to the isothermal-isobaric ensemble in Eq.~(\ref{NPT_ens_av_eq}) through MC simulations to obtain $\phi_S(\txtot, \mc, u_c^{\text{pot}})$, here approximated by Eq.~(\ref{eqn_phi_approx}). Construct the nucleation barrier $V_S(\txtot, \xc)$ in Eq.~(\ref{eq_final_barr}).
  \item Use the configurations $C$ as input in short-time MD simulations to obtain $\bm{D}_{\xc,\xc}$.
  \item Solve the Fokker-Planck equation (\ref{f_simple_evolution_eq}) in the stationary limit to obtain the nucleation rate prediction.
\end{enumerate}
Subsequently we offer more details to each of the steps.

\subsection{Lennard-Jones Model System}

We simulate the interaction between atoms through the long-ranged Lennard-Jones potential, expressed at a separating distance $r$ by%
\begin{equation*}
	\phi(r) = \left\{\begin{array}{l l}
		4\epsilon \bigl[\bigl(\frac{\sigma}{r}\bigr)^{12} - \bigl(\frac{\sigma}{r}\bigr)^6 \bigr]	&	\quad \text{if $0<r < r_\text{c}$}\,,			\\
		0																&	\quad \text{otherwise}\,.					\\
	\end{array}\right. 
\end{equation*}
The cutoff radius is $r_\text{c}=5\times\sigma$. Typical values of $\sigma$ and $\epsilon$ for a monoatomic Lennard-Jones system, made of argon-like particles, are $\sigma =3.822\times10^{-10}\,\text{m}$ and $\epsilon = 1.654\times10^{-21}\,\text{J}$, with a particle's mass $m = 6.690\times10^{-26}\,\text{kg}$. We also introduce the typical time-scale $\tau = \sqrt{\sigma^2 m /\epsilon}$ on which the atomistic dynamics takes place. We performed both, the MC and MD simulations with $\mtot=1000$ particles in the isobaric-isothermal ensemble. We used periodic boundary conditions in all directions. With $1000$ particles system size effects are negligible and we benefit from the fact that the system is small enough so that only one nucleus is formed in the system. 

\subsection{Clustering of Particles}
\label{Sec_clustering}

For our simulations we need to formulate the microscopic cluster criterion $\Pi_{\theta(\bar{\Gamma})}$ introduced in Sec.~\ref{Sec_statevar} to evaluate nucleus properties. We here rely on the extension\cite{Wolde:1998} of ten Wolde et al.\ of the most popular and simple scheme, the Stillinger-cluster definition where nearby atoms, within a distance smaller than a cutoff distance $r_c =1.5\sigma$ are considered part of the same cluster. Ten Wolde et al.\ further distinguish between particles with a liquid like environment and vapor like environment. A particle has a liquid like environemnt if it has at least $5$ neighbours as defined by the Stillinger-criterion. All liquid like particles which are connected are considered part of the same nucleus. The scheme by ten Wolde et al.\ is computationally cheap and has been shown by Wedekind et al.\ \cite{Wedekind:2007adee} to offer several advantages over the simple Stillinger definition. The developed nonequilibrium statistical mechanics does not indicate which cluster criterion should be used, but once the cluster criterion is fixed, the formulation of our coarse-graining procedure is independent of it.  

\subsection{MC Simulations}
\label{sec_MC}

To sample the entropic potential $\phi_S$ we exploit standard MC simulations\cite{Wedekind:2007adee} in the isobaric-isothermal ensemble with fixed number of particles $N$, pressure $P$ and temperature $T$. The formation of a nucleus in the low supersaturation regime of the Lennard-Jones gas is a rare-event and so we resample to an importance sampling technique to sample regions in phase space that contain a nucleus. We use the umbrella-sampling technique to bias the simulation with the bias potential
\begin{equation}
\begin{split}
	 \Pi_{\mc,\text{bias}} &= -\frac{1}{2} k \left(\Pi_{\mc} - \mc \right)^2
\end{split}
\end{equation} 
that favours the presence of a nucleus with size $\mc$ in the system. We use different umbrella windows corresponding to different choices of $\mc$. In particular, we take umbrella windows from cluster size $10$ to $140$ usually in steps of $10$. The bias strength is chosen as $k = 0.25$. The cluster size $140$ is in all simulations beyond the critical size $\mc^{*}$. In our MC simulations of the $NPT$ ensemble, we attempt at each step a volume changing move with probability $1/N$. If the volume changing move does not take place, each particle is moved with a probability of $5/N$ in this step. The maximum distance in any direction a particle can move is set to $0.4\sigma$. This ensures in all our simulations an average acceptance probability of volume moves and particle moves of $50\%$. To sample each umbrella window we first initialize homogeneously distributed particles in space so that their density coincides with the density of the metastable state at the temperature and supersaturation we are investigating, see also Table~\ref{TABLEPARAM}. Subsequently we perform an equilibration run of $10^7$ MC steps and then start averaging for $10^8$ MC steps to obtain the entropic potential. 
\begin{figure}[!htb]
\centering
\includegraphics[width=8.5cm]{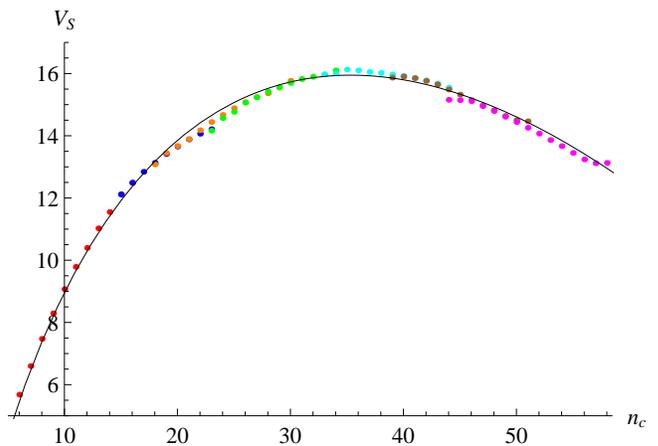}
\caption{Nucleation landscape $V_S(\txtot, \mc)$ for argon at temperature $T = 0.7\epsilon/\kb$ and supersaturation $S = 6.11$ near the critical size $\mc \approx 35$. Dots are obtained from Monte-Carlo simulations and the different colors indicate different umbrella windows. The solid line is a fit with Eq.~(\ref{fit_VS_eq}). }
\label{BarrierFig}
\end{figure}
The cluster criterion of ten Wolde et al.\ allows for a nucleus larger than $6$ particles only. We however fit the data obtained for $\phi_S$ between $\mc = 6$ and $140$ and assume this fit can be extrapolated down to $\mc = 1$. We found excellent fits of the nucleation landscape $V_S(\txtot, \xc)$ by Eq.~(\ref{eq_final_barr}) with
\begin{equation}
\begin{split}
V_S(\txtot, \mc) =&  -\log(S)(\mc - 1) + \eta_1 \left(\mc^{\frac{2}{3}}-1 \right) \\&
		     + \eta_2\left(\mc^{\frac{1}{3}} -1\right),
\label{fit_VS_eq}
\end{split}
\end{equation} 
where $\eta_1$ and $\eta_2$ depend on the temperature $T$ and supersaturation $S$. Furthermore, we fit the potential minimum $u_c^{\text{pot},0}(\mc) = e_0 \mc + e_s \mc^{2/3}$ and $H(\mc) = h_1\mc^{-3/2}$, see also Table~\ref{TABLEPARAM}. Here, $e_0$ can be interpreted as the bulk energy per particle of a nucleus, and $e_s$ as its corresponding surface energy. In Fig.~\ref{BarrierFig} we illustrate a typical nucleation landscape $V_S(\txtot, \mc)$ near the critical cluster size reconstructed from the MC simulations.
\renewcommand{\arraystretch}{1.5} 
\begin{table*}[ht] 
\caption{  Fit of the entropic potential (\ref{eq_final_barr}) and diffusion tensor (\ref{GK_eqeval}) with conventions and parameters introduced in Sec.~\ref{sec_MC} and Sec.~\ref{sec_MD}. In $e_0$ and $e_s$ as well as $h_1$ the error is smaller than the number of presented digits. The error in $D^0\tau$ has been discussed in Sec.~\ref{sec_MD}. Errors in $D_{\mc,\uc}^{r,0}\epsilon^2$ and $\xi^{r,0}m\epsilon$ are smaller than $0.05$ and errors in $D_{\uc,\uc}^{r,0}\epsilon^4$ smaller than $0.1$.}
\begin{adjustwidth}{1.9cm}{-1cm}
\begin{tabular*}{0.48\textwidth}{ccccccccccccc} 
$\kb T/\epsilon$			& 	
$S$ 					&  
$\rho_{g}\sigma^2$			&	
$\eta_1$					&	
$\eta_2$					&	
$e_0/\epsilon$				&	
$e_s/\epsilon $			&	 
$h_1\epsilon^2$ 		  			&	 
$D^0\tau$  			&	
$D_{\mc,\uc}^{r,0}\epsilon^2$  	&	 
$D_{\uc,\uc}^{r,0}\epsilon^4$  	&	 
$\xi^{r,0}m\epsilon$ \\   
\cline{1-13}
\cline{1-13}
$0.8$	&$2.98$	& $ 0.0200  $	&$7.81\pm0.04$	&$-9.73\pm0.25$	&	$-4.8 $&	$5.4$ 		&	 $6.8$	& 	$0.089$&   $0.72$&    $1.79$&    $0.77$	\\   
$0.8$	&$3.11$	& $ 0.0211  $	&$7.86\pm0.04$	&$-8.99 \pm0.12$&	$-4.7 $&	$5.3$ 		&	 $6.6$	& 	$0.132$&   $0.71$&    $1.50$&    $0.62$	\\   
$0.8$	&$3.33$	& $ 0.0230  $	&$7.64\pm0.05$	&$-7.73 \pm0.24$&	$-4.7 $&	$5.1$ 		&	 $6.6$	& 	$0.229$&   $0.68$&    $1.15$&    $0.50$ \\   
$0.8$	&$3.55$	& $ 0.0250  $	&$7.64\pm0.07$	&$-7.50\pm0.31$	&	$-4.6 $&	$4.9$ 		&	 $6.5$	& 	$0.314$&   $0.68$&    $0.88$&    $0.39$	\\   
$0.7$	&$5.30$	& $ 0.0106  $	&$10.62\pm0.07$&$-12.19\pm0.23$	&	$-4.9$	&	$5.1$ 		&	 $7.5$	& 	$0.054$&   $0.75$&    $1.42$&    $0.80$	\\   
$0.7$	&$6.11$	& $ 0.0125  $	&$10.27\pm0.05$&$-9.27 \pm0.38$	&	$-4.8$	&	$4.9$ 		&	 $8.5$	& 	$0.145$&   $0.74$&    $1.16$&    $0.37$	\\   
$0.6$	&$16.90$ &$ 0.0080  $	&$12.21\pm0.06$&$-7.63 \pm0.70$	&	$-5.3$	&	$5.7$ 		&	 $4.8$	& 	$0.056$&   $0.54$&    $0.89$&    $0.49$	\\   
$0.6$	&$15.60$ &$ 0.0073  $	&$12.04\pm0.07$&$-7.26 \pm0.21$	&	$-5.3$	&	$5.8$ 		&	 $5.1$	& 	$0.038$&   $0.61$&    $1.03$&    $0.63$ \\   
$0.6$	&$14.00$ &$ 0.0065  $	&$11.89\pm0.05$&$ -6.04\pm0.18$	&	$-5.3$	&	$5.8$ 		&	 $5.3$	& 	$0.032$&   $0.68$&    $1.20$&    $0.77$ \\   
$0.6$	&$11.95$ &$ 0.005   $	&$12.06\pm0.08$&$-7.19 \pm0.12$	&	$-5.3$	&	$5.7$ 		&	 $6.4$	& 	$0.025$&   $0.74$&    $1.42$&    $0.81$	\\   
$0.5$	&$39.37$ &$ 0.0021  $	&$16.34\pm0.09$&$-11.08 \pm0.32$&	$-5.6$	&	$6.3$ 		&	 $5.1$	& 	$0.013$&   $0.70$&    $1.00$&    $0.62$	\\   
$0.5$	&$43.31$ &$ 0.0023  $	&$16.57\pm0.10$&$-11.94\pm0.28$	&	$-5.6$	&	$6.2$ 		&	 $5.2$	& 	$0.012$&   $0.67$&    $0.85$&    $0.56$	\\   
$0.5$	&$48.70$ &$ 0.0026  $	&$16.74\pm0.09$&$-12.53\pm0.23$	&	$-5.6$	&	$6.3$ 		&	 $5.4$	& 	$0.012$&   $0.58$&    $0.69$&    $0.48$	\\     
\end{tabular*}\end{adjustwidth}
\label{TABLEPARAM}
\end{table*}%

\subsection{MD Simulations}
\label{sec_MD}

The atomic positional configurations harvested from MC simulations are used as initial conditions for our short-time molecular dynamics simulations in the microcanonical ensembles. We can equally use the isobaric-isothermal ensemble at temperature $T$ and pressure $P$ to sample the short-time correlations of fluctuations determining the diffusion coefficient since on these short time-scales the macroscopic state of the system cannot be affected. Microcanonical simulations are more robust since there is no thermostat that can affect the nucleus state. If the isobaric-isothermal ensemble is used, the thermostats corresponding to $T$ and $P$ must be ensured to produce system relaxation times of the thermostated variables that drop off much slower than the time-scale on which relevant microscopic correlations drop off. Otherwise, the nucleus momentum and energy can be affected through the thermostat on the same time-scale as microscopic collisions of atoms of the ambient phase and the nucleus take place, resulting then in erroneous estimates of the diffusion tensor\cite{Wedekind:2007adee}. For simplicity we have sampled the initial momentum of all atoms in our MD simulations with a Gaussian distribution at the temperature $T$ we are investigating. In this work we omit studying the energy dependence of the diffusion tensor and solely capture its size dependence. To sample the diffusion tensor at the critical cluster size for any given temperature and supersaturation regime, we take $32$ configurations harvested from MC runs that contain a nucleus of the critical size and then run the MD simulation for a total time of $25\tau$ around five times larger than the collision time-scale $ \tau_{\text{GK}} \approx 5\tau $ of gas particles with the nucleus. By averaging the dynamics of all $32$ simulations we are able to obtain the required mean-square displacements (\ref{GK_eq}). The corresponding correlation functions $ \len \Delta_{\tau_{\text{GK}} } \Pi_{\xc}\Delta_{\tau_{\text{GK}} } \Pi_{\xc} \ren_{(\xtot, \xc)}$ are illustrated in Fig.~\ref{Figlocequilibriutemp}, and within the collision time-scale a linear regime is recognizable from which the diffusion coefficient can be obtained.
\begin{figure}[!htb]
\centering
\includegraphics[width=8.5cm]{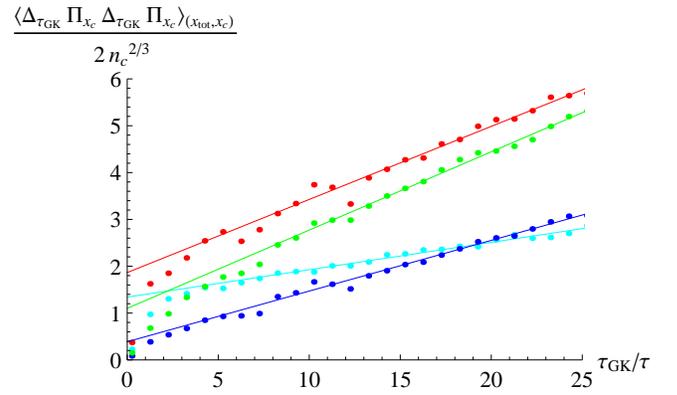}
\caption{Diffusion matrix normalized by $\mc^{2/3}$ for nucleation of argon at temperature $T$ and supersaturation $S=6.11$. The dots denote the mean-square displacement of $\mc$ with $\mc$ (red), between $\mc$ with $\uc$ (blue), between $\uc$ with $\uc$ and $\pc^{x}$ with $\pc^{x}$, where $\pc^{x}$ is the momentum in $x$-direction. The solid lines, the slope of which enters the prediction of the diffusion tensor, are fits to the linear regime of these correlations.}
\label{Figlocequilibriutemp}
\end{figure}
In Table~\ref{TABLEPARAM} we have listed all obtained diffusion coefficients with the convention
\begin{equation}
\begin{split} 
\bm{D}(\xtot, \xc) =    
 \begin{pmatrix}
  \xi^{r,0} \mathbf{I}_{3\times 3}        & 0  \\
  0   &1 &  D_{\mc,\uc}^{r,0}       \\ 
  0   & D_{\mc,\uc}^{r,0} &  D_{\uc,\uc}^{r,0}       \\ 
 \end{pmatrix}
D^{0} \mc^{2/3}.
\label{GK_eqeval}
\end{split}
\end{equation} 
We also evaluated the diffusion coefficients for the temperature $T=0.7$ and supersaturation ratio $6.11$ for nucleus sizes distinct from the critical size and realized that they are well approximated by  Eq.~(\ref{GK_eqeval}) for nuclei down to $15$ atoms. The scaling (\ref{GK_eqeval}) was also valid in the post-critical regime. We hence assume the validity of Eq.~(\ref{GK_eqeval}) for all nucleus sizes. In his work\cite{Barrett:1994,Barrett:2008}, Barret gives explicit expressions for the coefficients $D_{\mc,\uc}^{r,0}$ and $D_{\uc,\uc}^{r,0} $ derived by kinetic theory arguments, see also Sec.~\ref{sec_Argon_MFKT}. We did not find his theoretical predictions to match with our computer experiments. But we found rather good agreement of his prediction of $D^{0} $ with ours.

In evaluating the coefficients $ \len \Delta_{\tau_{\text{GK}}} \Pi_{\xc}\Delta_{\tau_{\text{GK}}} \Pi_{\xc} \ren_{(\xtot, \xc)}$ we notice that the microscopic process behind its contribution is the exchange of mass, momentum and energy of the nucleus with the ambient phase. We adopted the cluster criterion of ten Wolde et al.\ in our MC simulations, and within this criterion there are always particles located at the surface of the nucleus which are not part of the nucleus since they have less than $5$ neighbours. These particles can however stay at the surface of the nucleus for a considerable time-scale much larger than the collision time-scale $\tau_{\text{GK}}$ and even rapidly oscillate between being part of the nucleus and part of the ambient phase. These oscillations lead to significant noise effects that make it difficult to evaluate the true diffusion coefficients, which by nature should only be affected by particles that truly collide, are absorbed or emitted by the nucleus and do not fall in the undesirable class of oscillating surface particles, an inherent drawback of the ten Wolde cluster criterion. The contribution to the diffusion coefficient from the oscillating particles should dissappear if we average over sufficiently many simulations. The resulting noise effects are understood by the fact that the number of attempts with which the outer shell particles oscillate between being a nucleus property and a gas property outnumbers the true collision attempts from the ambient phase. We found that we can significantly improve the statistics of $ \len \Delta_{\tau_{\text{GK}}} \Pi_{\xc}\Delta_{\tau_{\text{GK}}} \Pi_{\xc} \ren_{(\xtot, \xc)}$ by using instead of the ten Wolde cluster criterion, the Stillinger-cluster criterion only so that now the oscillating particles in the ten Wolde cluster definition belong to the nucleus as well and thereby, the nucleus state $\xc$ only changes through desirable collision processes. It is expected that this procedure results in an error in the obtained diffusion coefficients since by deciding that the outer layers of the nucleus is now also a nucleus property, the original nucleus state obtained from the ten Wolde definition is modified. This error only affects the prefactor $D_{\mc,\mc}^{0} $ in Eq.~(\ref{GK_eqeval}). The Stillinger criterion gives typically nucleus sizes $\mc^{\text{Stillinger}}$ that are $3/2$ larger than the ten-Wolde nucleus size $\mc^{\text{ten Wolde}}$. In Eq.~(\ref{GK_eqeval}) we set $D^{0} = \bm{D}_{\mc,\mc}/\mc^{2/3}$. Suppose we obtained $ \bm{D}_{\mc,\mc}$ in our simulation around the critical size $\mc^{*,\text{Stillinger}} \approx 3/2 \mc^{*,\text{Wolde}} $ with the Stillinger criterion, then $D^{0} = \bm{D}_{\mc,\mc}/(\mc^{\text{Stillinger}})^{2/3}$ and if we use the ten-Wolde definition, we must get the same coefficient $\bm{D}_{\mc,\mc}$, but $D^{0} = \bm{D}_{\mc,\mc}/(\mc^{\text{Wolde}})^{2/3}$. This means that the two values obtained for $D^{0} $ are distinct by a factor $(\mc^{*,\text{Stillinger}}/\mc^{*,\text{ten Wolde}})^{3/2} \approx 2$, so that the nucleation rate prediction that depends linearly on $D^0$ is affected by an error of a factor of two. In nucleation theory such errors are completely negligible.

\subsection{Rate-Estimation}

The rate prediction for the truncated theory using $\xc^{\text{trc}}= (\mc)$ as state variable is given by the classical formula
\begin{equation}
\begin{split} 
J_{\text{trc}} =& \sqrt{-\frac{1}{2\pi}\left.\frac{\partial^2 V_S(\txtot,\mc)}{\partial \mc^2} \right|_{\mc^{*}}} \rho_{g}   
		\\& \times D^{0}\left(\mc^{*}\right)^{\frac{3}{2}}  e^{- \Delta V_S(\txtot,\mc^{*}) },
\label{RATE_EST_AMP}
\end{split}
\end{equation}
where $\mc^{*}$ is the critical size of the truncated nucleation landscape $V_S(\txtot,\mc^{*})$ and $ \Delta V_S(\txtot,\mc^{*})$ the barrier height, see also Eq.~(\ref{fit_VS_eq}).
In order to obtain the nucleation rate of the full theory $J_{\text{ext}}$, we performed Brownian dynamics (BD) simulations of the full Fokker-Planck equation (\ref{f_simple_evolution_eq}) and the Fokker-Planck equation of the truncated theory (\ref{Zeldovic_eq}) to obtain the correction to the nucleation rate (\ref{RATE_EST_AMP}) through nonisothermal effects and Brownian motion. For the BD simulation we closely follow earlier work\cite{Schweizer:2014}. In Table~\ref{tab_RATE_PREDICTION} we tabulated the different obtained rates and corrections $\alpha_{\text{non-iso}}$ and $\alpha_{\text{BM}}$ to $J_{\text{trc}}$ through nonisothermal and Brownian effects of the nucleus. We roughly estimated these correction by first performing the BD simulations in an isothermal setup, but with the correction for Brownian motion activated to obtain $\alpha_{\text{BM}}$ and subsequently we performed BD simulations incorporating all state variables and the additional correction was attributed to $\alpha_{\text{non-iso}}$.
\renewcommand{\arraystretch}{1.5} 
\begin{table*}[ht] 
\caption{\label{tab_RATE_PREDICTION} Nucleation rate predictions for a metastable Lennard-Jones gas at various temperatures $T$ and supersaturations $S$. $J_{\text{MD}}$ is the prediction of MD simulations\cite{Diemand:2013} (an asterix on top of the data indicates that the rate is obtained through the universal scaling relation\cite{Tanaka:2014scaling}). $J_{\text{trc}}$ is the prediction of the standard technique using the truncated state space and $J_{\text{ext}}$ uses the full state space. $\alpha_{\text{non-iso}}$ and $\alpha_{\text{BM}}$ estimates the correction of the nucleation rate $J_{\text{trc}}$ due to nonisothermal effects and Brownian motion of the nucleus respectively.   }
\begin{adjustwidth}{2.5cm}{-1cm}
\begin{tabular*}{0.48\textwidth}{cccccccc} 
$\kb T/\epsilon$				& 	
$S$ 									&	
$\log_{10}\left(\frac{J_{\text{MD}}}{\sigma^2 \tau}\right)$ 		&
$\log_{10}\left(\frac{J_{\text{trc}}}{\sigma^2 \tau}\right)$  		&   
$\log_{10}\left(\frac{J_{\text{ext}}}{\sigma^2 \tau}\right)$ 				&   
$\log_{10}\left(\frac{\alpha_{\text{non-iso}}}{\sigma^2 \tau}\right)$ 			&   
$\log_{10}\left(\frac{\alpha_{\text{BM}}}{\sigma^2 \tau}\right)$ 			\\   
\cline{1-7}
\cline{1-7}
$0.8$		&	$2.98$	&	$-16\pm0.5$		&	$-11.43\pm0.29 $	&	$-16.08\pm0.61 $	&	$-1.0 $	&$-3.7 $\\ 
$0.8$		&	$3.11$	&	$-15.53^{*}$		&	$-11.33\pm0.20 $	&	$-16.03\pm0.42 $ &	$-1.0 $	&$-3.6 $\\  
$0.8$		&	$3.33$	&	$-13.47\pm0.07$		&	$-9.71  \pm0.27 $	&	$-14.34\pm0.58 $ &	$-1.2 $	&$-3.4 $\\  
$0.8$		&	$3.55$	&	$-11.90\pm0.02$		&	$-8.59   \pm0.33$	&	$-13.20\pm0.60 $ &	$-1.4 $	&$-3.2 $\\  
$0.7$		&	$5.30$	&	$-16.24^{*}$		&	$-11.89 \pm0.31$	&	$-16.40\pm0.59 $ &	$-1.2 $	&$-3.3 $\\  
$0.7$		&	$6.11$	&	$-13.28^{*}$		&	$-10.43 \pm0.30$	&	$-14.83\pm0.68 $	&	$-1.3 $	&$-3.1 $\\  
$0.6$		&	$16.90$ &	$-11.96\pm0.01$		&	$-8.52  \pm0.30 $	&	$-12.21\pm0.62 $	&	$-1.4 $	&$-2.3 $\\  
$0.6$		&	$15.60$ &	$-12.81\pm0.04$		&	$-9.11  \pm0.16 $ 	&	$-12.86\pm0.35 $	&	$-1.3 $	&$-2.5 $\\   
$0.6$		&	$14.00$ &	$-14.58\pm0.07$		&	$-10.53 \pm0.14$	&	$-14.39\pm0.29 $	&	$-1.3 $	&$-2.6 $\\  
$0.6$		&	$11.95$ &	$-16.45\pm0.98$		&	$-11.83\pm0.18 $	&	$-15.88\pm0.36 $	&	$-1.2 $	&$-2.8 $\\   
$0.5$		&	$39.37$ &	$-16.77^{*}$		&	$-11.49 \pm0.22$	&	$-15.43\pm0.48 $	&	$-1.5 $	&$-2.4 $\\   
$0.5$		&	$43.31$ &	$-15.89^{*}$		&	$-10.80\pm Table0.21 $	&	$-14.82\pm0.40 $	&	$-1.7 $	&$-2.3 $\\    
$0.5$		&	$48.70$ &	$-15.27\pm0.05$		&	$-10.03 \pm0.17$	&	$-13.97\pm0.34 $	&	$-1.7 $	&$-2.5 $\\       
\end{tabular*}\end{adjustwidth}
\end{table*}%
\begin{figure}[!htb]
\centering
\includegraphics[width=8.5cm]{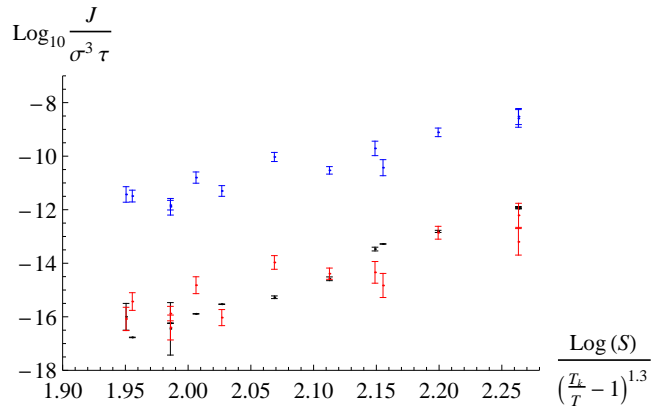}
\caption{Comparison of argon nucleation rate predictions of the standard Monte-Carlo technique (blue) with predictions of our theory (red) where the nuclei are described through an extended state space. Also shown are the results of brute-force molecular dynamics simulations\cite{Diemand:2013} (black). Dotted lines are a guide to the eyes. }
\label{rateplotFig}
\end{figure}
We observed that in lowering the ambient temperature $T$ the nonisothermal effects become more pronounced and comparable with the Brownian effects. In the high temperature regime however, Brownian effects dominate. The lower the temperature the more latent heat is exchanged between the nucleus and ambient phase so that we expect this behavior. In Fig~\ref{rateplotFig} we compare the predictions $J_{\text{ext}}$ with the brute-force MD rates $J_{\text{MD}}$ obtained by Diemand et al.\ \cite{Diemand:2013} and plotted it in a manner that highlights the intrinsic universal scaling behavior suggested by Tanaka et al.\ \cite{Tanaka:2014scaling}, where the MD nucleation rate is recognized as a universal function of $\log(S)/(T_k/T -1)^{1.3}$ with $T_k = 1.312\epsilon/\kb$ the critical temperature. The rates obtained by the standard truncated theory deviate by $3-5$ orders of magnitude from the exact MD rates whereas the rates obtained within our extended theory yields a quite excellent agreement.

\section{Conclusion and Outlook}

We have derived the nonequilibrium statistical mechanics of nucleation with an extended state space within the GENERIC coarse-graining technique. We have shown that the truncated state space where only the number of atoms in a nucleus is a relevant variable leads to systematically erronoreous prediction of the nucleation rate of $3-5$ orders of magnitude for the prototypical studied nucleation of liquid drops in a supersaturated Lennard-Jones gas. When we extend the nuclei degree of freedom to involve in addition its internal energy and momentum, this lack in prediction can be significantly improved. For the example of the MFKT of nucleation we also showed that truncated theories can easily be extended to a richer state space to improve predictions. Within our approach we were rigorously able to derive the correction of the nucleation landscape due to energy fluctuations and momentum fluctuations, and in both cases we were able to reproduce the up to date results. This includes the non-Gaussian temperature fluctuations inherent for small nuclei following the law of McGraw and Laviolette, and the mesoscopic irreversible thermodynamics contribution of Reguera and Rubi for Brownian motion. The latter, as we found, must be corrected by a logarithmic contribution in the cluster size that arrises from normalization reasons, and can probabily not be obtained by bare thermodynamic argumentations as done by Reguera and Rubi. Within our statistical mechanics approach we were also able to state the proper definition of the nucleus temperature that leads to the temperature fluctuations predicted by McGraw and Laviolette. 

We predicted the nucleation rate in a metastable Lennard-Jones gas in the experimentally relevant low supersaturation regime and found excellent agreement with recent large-scale brute-force MD simulations. As argued by Diemand et al.\, when the nuclei size is smaller than about $16$ atoms, nuclei tend to spin so that the angular momentum would enter as an additional important state variable. In the majority of our simulations the critical size was well beyond $16$ so that these corrections are irrelevant. From this point of view it might be interesting to include additional state variables into the description such as the angular momentum or the shape of the nucleus. In this respect our coarse-graining procedure could also be extended to multicomponent system for more realistic applications. In our work we relied on a cluster definition to group the atoms as members of the gas phase or the nucleus of interest. Some work has been done avoiding the introduction of an artificial cluster criterion and we also expect useful extensions of our work in this directions, possibly following the lines of Kusaka et al.\ \cite{Kusaka:1998b, Kusaka:1999oparam}.

\section{Acknowledgments}

We thank Hans Christian \"Ottinger for various hints and ideas to improve this work significantly. Table

\bibliography{bibTex}

\begin{thebibliography}{59}%
\makeatletter
\providecommand \@ifxundefined [1]{%
 \@ifx{#1\undefined}
}%
\providecommand \@ifnum [1]{%
 \ifnum #1\expandafter \@firstoftwo
 \else \expandafter \@secondoftwo
 \fi
}%
\providecommand \@ifx [1]{%
 \ifx #1\expandafter \@firstoftwo
 \else \expandafter \@secondoftwo
 \fi
}%
\providecommand \natexlab [1]{#1}%
\providecommand \enquote  [1]{``#1''}%
\providecommand \bibnamefont  [1]{#1}%
\providecommand \bibfnamefont [1]{#1}%
\providecommand \citenamefont [1]{#1}%
\providecommand \href@noop [0]{\@secondoftwo}%
\providecommand \href [0]{\begingroup \@sanitize@url \@href}%
\providecommand \@href[1]{\@@startlink{#1}\@@href}%
\providecommand \@@href[1]{\endgroup#1\@@endlink}%
\providecommand \@sanitize@url [0]{\catcode `\\12\catcode `\$12\catcode
  `\&12\catcode `\#12\catcode `\^12\catcode `\_12\catcode `\%12\relax}%
\providecommand \@@startlink[1]{}%
\providecommand \@@endlink[0]{}%
\providecommand \url  [0]{\begingroup\@sanitize@url \@url }%
\providecommand \@url [1]{\endgroup\@href {#1}{\urlprefix }}%
\providecommand \urlprefix  [0]{URL }%
\providecommand \Eprint [0]{\href }%
\providecommand \doibase [0]{http://dx.doi.org/}%
\providecommand \selectlanguage [0]{\@gobble}%
\providecommand \bibinfo  [0]{\@secondoftwo}%
\providecommand \bibfield  [0]{\@secondoftwo}%
\providecommand \translation [1]{[#1]}%
\providecommand \BibitemOpen [0]{}%
\providecommand \bibitemStop [0]{}%
\providecommand \bibitemNoStop [0]{.\EOS\space}%
\providecommand \EOS [0]{\spacefactor3000\relax}%
\providecommand \BibitemShut  [1]{}%
\let\auto@bib@innerbib\@empty
\bibitem [{\citenamefont {Yang}\ \emph {et~al.}(2006)\citenamefont {Yang},
  \citenamefont {Deymier}, \citenamefont {Wang}, \citenamefont {Guzman},\ and\
  \citenamefont {Hoying}}]{Yang:2006}%
  \BibitemOpen
  \bibfield  {author} {\bibinfo {author} {\bibfnamefont {Y.}~\bibnamefont
  {Yang}}, \bibinfo {author} {\bibfnamefont {P.~A.}\ \bibnamefont {Deymier}},
  \bibinfo {author} {\bibfnamefont {L.}~\bibnamefont {Wang}}, \bibinfo {author}
  {\bibfnamefont {R.}~\bibnamefont {Guzman}}, \ and\ \bibinfo {author}
  {\bibfnamefont {J.~B.}\ \bibnamefont {Hoying}},\ }\href@noop {} {\bibfield
  {journal} {\bibinfo  {journal} {Biotechnol. Prog.}\ }\textbf {\bibinfo
  {volume} {22}},\ \bibinfo {pages} {303--312} (\bibinfo {year}
  {2006})}\BibitemShut {NoStop}%
\bibitem [{\citenamefont {Longuet}\ \emph {et~al.}(2014)\citenamefont
  {Longuet}, \citenamefont {Yamada}, \citenamefont {Chen}, \citenamefont
  {Baigl},\ and\ \citenamefont {Fattaccioli}}]{Longuet:2014}%
  \BibitemOpen
  \bibfield  {author} {\bibinfo {author} {\bibfnamefont {C.}~\bibnamefont
  {Longuet}}, \bibinfo {author} {\bibfnamefont {A.}~\bibnamefont {Yamada}},
  \bibinfo {author} {\bibfnamefont {Y.}~\bibnamefont {Chen}}, \bibinfo {author}
  {\bibfnamefont {D.}~\bibnamefont {Baigl}}, \ and\ \bibinfo {author}
  {\bibfnamefont {J.}~\bibnamefont {Fattaccioli}},\ }\href@noop {} {\bibfield
  {journal} {\bibinfo  {journal} {J. Chem. Phys.}\ }\textbf {\bibinfo {volume}
  {386}},\ \bibinfo {pages} {179--182} (\bibinfo {year} {2014})}\BibitemShut
  {NoStop}%
\bibitem [{\citenamefont {ter Horst}, \citenamefont {Kramer},\ and\
  \citenamefont {Jansens}(2002)}]{Horst:2002aa}%
  \BibitemOpen
  \bibfield  {author} {\bibinfo {author} {\bibfnamefont {H.}~\bibnamefont {ter
  Horst}}, \bibinfo {author} {\bibfnamefont {J.~H.~M.}\ \bibnamefont {Kramer}},
  \ and\ \bibinfo {author} {\bibfnamefont {P.~J.}\ \bibnamefont {Jansens}},\
  }\href@noop {} {\bibfield  {journal} {\bibinfo  {journal} {Cryst. Growth
  Des.}\ }\textbf {\bibinfo {volume} {2(5)}},\ \bibinfo {pages} {351} (\bibinfo
  {year} {2002})}\BibitemShut {NoStop}%
\bibitem [{\citenamefont {Kuba}\ and\ \citenamefont
  {Murakami}(2010)}]{Kuba:2010}%
  \BibitemOpen
  \bibfield  {author} {\bibinfo {author} {\bibfnamefont {N.}~\bibnamefont
  {Kuba}}\ and\ \bibinfo {author} {\bibfnamefont {M.}~\bibnamefont
  {Murakami}},\ }\href@noop {} {\bibfield  {journal} {\bibinfo  {journal}
  {Atmos. Chem. Phys.}\ }\textbf {\bibinfo {volume} {10}},\ \bibinfo {pages}
  {3335–3351} (\bibinfo {year} {2010})}\BibitemShut {NoStop}%
\bibitem [{\citenamefont {Krugel}(2003)}]{KrugelB2003}%
  \BibitemOpen
  \bibfield  {author} {\bibinfo {author} {\bibfnamefont {E.}~\bibnamefont
  {Krugel}},\ }\href@noop {} {\emph {\bibinfo {title} {The Physics of
  Interstellar Dust (Series in Astronomy and Astrophysics)}}}\ (\bibinfo
  {publisher} {Institute of Physics Publishing},\ \bibinfo {year}
  {2003})\BibitemShut {NoStop}%
\bibitem [{\citenamefont {Diemand}\ \emph {et~al.}(2013)\citenamefont
  {Diemand}, \citenamefont {Ang\'{e}lil}, \citenamefont {Tanaka},\ and\
  \citenamefont {Tanaka}}]{Diemand:2013}%
  \BibitemOpen
  \bibfield  {author} {\bibinfo {author} {\bibfnamefont {J.}~\bibnamefont
  {Diemand}}, \bibinfo {author} {\bibfnamefont {R.}~\bibnamefont
  {Ang\'{e}lil}}, \bibinfo {author} {\bibfnamefont {K.~K.}\ \bibnamefont
  {Tanaka}}, \ and\ \bibinfo {author} {\bibfnamefont {H.}~\bibnamefont
  {Tanaka}},\ }\href@noop {} {\bibfield  {journal} {\bibinfo  {journal} {J.
  Chem. Phys.}\ }\textbf {\bibinfo {volume} {139}},\ \bibinfo {pages} {074309}
  (\bibinfo {year} {2013})}\BibitemShut {NoStop}%
\bibitem [{\citenamefont {Chkonia}\ \emph {et~al.}(2009)\citenamefont
  {Chkonia}, \citenamefont {W{\"o}lk}, \citenamefont {Strey}, \citenamefont
  {Wedekind},\ and\ \citenamefont {Reguera}}]{Chkonia:2009}%
  \BibitemOpen
  \bibfield  {author} {\bibinfo {author} {\bibfnamefont {G.}~\bibnamefont
  {Chkonia}}, \bibinfo {author} {\bibfnamefont {J.}~\bibnamefont {W{\"o}lk}},
  \bibinfo {author} {\bibfnamefont {R.~S.}\ \bibnamefont {Strey}}, \bibinfo
  {author} {\bibfnamefont {J.}~\bibnamefont {Wedekind}}, \ and\ \bibinfo
  {author} {\bibfnamefont {D.}~\bibnamefont {Reguera}},\ }\href@noop {}
  {\bibfield  {journal} {\bibinfo  {journal} {J. Chem. Phys.}\ }\textbf
  {\bibinfo {volume} {130}},\ \bibinfo {pages} {064505} (\bibinfo {year}
  {2009})}\BibitemShut {NoStop}%
\bibitem [{\citenamefont {Kraska}(2006)}]{Kraska:2006}%
  \BibitemOpen
  \bibfield  {author} {\bibinfo {author} {\bibfnamefont {T.}~\bibnamefont
  {Kraska}},\ }\href@noop {} {\bibfield  {journal} {\bibinfo  {journal} {J.
  Chem. Phys.}\ }\textbf {\bibinfo {volume} {124}},\ \bibinfo {pages} {054507}
  (\bibinfo {year} {2006})}\BibitemShut {NoStop}%
\bibitem [{\citenamefont {Tanaka}\ \emph {et~al.}(2005)\citenamefont {Tanaka},
  \citenamefont {Kawamura}, \citenamefont {Tanaka},\ and\ \citenamefont
  {Nakazawa}}]{Tanaka:2005}%
  \BibitemOpen
  \bibfield  {author} {\bibinfo {author} {\bibfnamefont {K.~K.}\ \bibnamefont
  {Tanaka}}, \bibinfo {author} {\bibfnamefont {K.}~\bibnamefont {Kawamura}},
  \bibinfo {author} {\bibfnamefont {H.}~\bibnamefont {Tanaka}}, \ and\ \bibinfo
  {author} {\bibfnamefont {K.}~\bibnamefont {Nakazawa}},\ }\href@noop {}
  {\bibfield  {journal} {\bibinfo  {journal} {J. Chem. Phys.}\ }\textbf
  {\bibinfo {volume} {122}},\ \bibinfo {pages} {184514} (\bibinfo {year}
  {2005})}\BibitemShut {NoStop}%
\bibitem [{\citenamefont {Tanaka}\ \emph {et~al.}(2011)\citenamefont {Tanaka},
  \citenamefont {Tanaka}, \citenamefont {Yamamoto},\ and\ \citenamefont
  {Kawamura}}]{Tanaka:2011}%
  \BibitemOpen
  \bibfield  {author} {\bibinfo {author} {\bibfnamefont {K.~K.}\ \bibnamefont
  {Tanaka}}, \bibinfo {author} {\bibfnamefont {H.}~\bibnamefont {Tanaka}},
  \bibinfo {author} {\bibfnamefont {T.}~\bibnamefont {Yamamoto}}, \ and\
  \bibinfo {author} {\bibfnamefont {K.}~\bibnamefont {Kawamura}},\ }\href@noop
  {} {\bibfield  {journal} {\bibinfo  {journal} {J. Chem. Phys.}\ }\textbf
  {\bibinfo {volume} {134}},\ \bibinfo {pages} {204313} (\bibinfo {year}
  {2011})}\BibitemShut {NoStop}%
\bibitem [{\citenamefont {Wedekind}\ \emph {et~al.}(2007)\citenamefont
  {Wedekind}, \citenamefont {W{\"o}lk}, \citenamefont {Reguera},\ and\
  \citenamefont {Strey}}]{Wedekind:2007}%
  \BibitemOpen
  \bibfield  {author} {\bibinfo {author} {\bibfnamefont {J.}~\bibnamefont
  {Wedekind}}, \bibinfo {author} {\bibfnamefont {J.}~\bibnamefont {W{\"o}lk}},
  \bibinfo {author} {\bibfnamefont {D.}~\bibnamefont {Reguera}}, \ and\
  \bibinfo {author} {\bibfnamefont {R.}~\bibnamefont {Strey}},\ }\href@noop {}
  {\bibfield  {journal} {\bibinfo  {journal} {J. Chem. Phys.}\ }\textbf
  {\bibinfo {volume} {127}},\ \bibinfo {pages} {154515} (\bibinfo {year}
  {2007})}\BibitemShut {NoStop}%
\bibitem [{\citenamefont {Yasuoka}\ and\ \citenamefont
  {Matsumoto}(1998)}]{Yasuoka:1998}%
  \BibitemOpen
  \bibfield  {author} {\bibinfo {author} {\bibfnamefont {K.}~\bibnamefont
  {Yasuoka}}\ and\ \bibinfo {author} {\bibfnamefont {M.}~\bibnamefont
  {Matsumoto}},\ }\href@noop {} {\bibfield  {journal} {\bibinfo  {journal} {J.
  Chem. Phys.}\ }\textbf {\bibinfo {volume} {109}},\ \bibinfo {pages} {8451}
  (\bibinfo {year} {1998})}\BibitemShut {NoStop}%
\bibitem [{\citenamefont {Kalikmanov}(2013)}]{KalikmanovB2013}%
  \BibitemOpen
  \bibfield  {author} {\bibinfo {author} {\bibfnamefont {V.}~\bibnamefont
  {Kalikmanov}},\ }\href@noop {} {\emph {\bibinfo {title} {Nucleation
  Theory}}}\ (\bibinfo  {publisher} {Springer},\ \bibinfo {year}
  {2013})\BibitemShut {NoStop}%
\bibitem [{\citenamefont {Girshick}\ and\ \citenamefont
  {Chiu}(1990)}]{Girshick:1990}%
  \BibitemOpen
  \bibfield  {author} {\bibinfo {author} {\bibfnamefont {S.}~\bibnamefont
  {Girshick}}\ and\ \bibinfo {author} {\bibfnamefont {C.-P.}\ \bibnamefont
  {Chiu}},\ }\href@noop {} {\bibfield  {journal} {\bibinfo  {journal} {J. Chem.
  Phys.}\ }\textbf {\bibinfo {volume} {93}},\ \bibinfo {pages} {1273} (\bibinfo
  {year} {1990})}\BibitemShut {NoStop}%
\bibitem [{\citenamefont {Dillmann}\ and\ \citenamefont
  {Meier}(1991)}]{Dillmann:1991}%
  \BibitemOpen
  \bibfield  {author} {\bibinfo {author} {\bibfnamefont {A.}~\bibnamefont
  {Dillmann}}\ and\ \bibinfo {author} {\bibfnamefont {G.~E.~A.}\ \bibnamefont
  {Meier}},\ }\href@noop {} {\bibfield  {journal} {\bibinfo  {journal} {J.
  Chem. Phys.}\ }\textbf {\bibinfo {volume} {94}},\ \bibinfo {pages} {3872}
  (\bibinfo {year} {1991})}\BibitemShut {NoStop}%
\bibitem [{\citenamefont {Delale}\ and\ \citenamefont
  {Meier}(1993)}]{Delale:1991}%
  \BibitemOpen
  \bibfield  {author} {\bibinfo {author} {\bibfnamefont {C.~F.}\ \bibnamefont
  {Delale}}\ and\ \bibinfo {author} {\bibfnamefont {G.~E.~A.}\ \bibnamefont
  {Meier}},\ }\href@noop {} {\bibfield  {journal} {\bibinfo  {journal} {J.
  Chem. Phys.}\ }\textbf {\bibinfo {volume} {98}},\ \bibinfo {pages} {9850}
  (\bibinfo {year} {1993})}\BibitemShut {NoStop}%
\bibitem [{\citenamefont {Kalikmanov}\ and\ \citenamefont {van
  Dongen}(1995)}]{Kalikmanov:1995}%
  \BibitemOpen
  \bibfield  {author} {\bibinfo {author} {\bibfnamefont {V.~I.}\ \bibnamefont
  {Kalikmanov}}\ and\ \bibinfo {author} {\bibfnamefont {M.~E.~H.}\ \bibnamefont
  {van Dongen}},\ }\href@noop {} {\bibfield  {journal} {\bibinfo  {journal} {J.
  Chem. Phys.}\ }\textbf {\bibinfo {volume} {103}},\ \bibinfo {pages} {4250}
  (\bibinfo {year} {1995})}\BibitemShut {NoStop}%
\bibitem [{\citenamefont {Laaksonen}, \citenamefont {Ford},\ and\ \citenamefont
  {Kulmala}(1994)}]{Laaksonen:1994}%
  \BibitemOpen
  \bibfield  {author} {\bibinfo {author} {\bibfnamefont {A.}~\bibnamefont
  {Laaksonen}}, \bibinfo {author} {\bibfnamefont {I.}~\bibnamefont {Ford}}, \
  and\ \bibinfo {author} {\bibfnamefont {M.}~\bibnamefont {Kulmala}},\
  }\href@noop {} {\bibfield  {journal} {\bibinfo  {journal} {Phys. Rev. E}\
  }\textbf {\bibinfo {volume} {49}},\ \bibinfo {pages} {5517} (\bibinfo {year}
  {1994})}\BibitemShut {NoStop}%
\bibitem [{\citenamefont {Reguera}\ and\ \citenamefont
  {Reiss}(2004)}]{Reguera:2004}%
  \BibitemOpen
  \bibfield  {author} {\bibinfo {author} {\bibfnamefont {D.}~\bibnamefont
  {Reguera}}\ and\ \bibinfo {author} {\bibfnamefont {H.}~\bibnamefont
  {Reiss}},\ }\href@noop {} {\bibfield  {journal} {\bibinfo  {journal} {J.
  Phys. Chem. B}\ }\textbf {\bibinfo {volume} {108 (51)}},\ \bibinfo {pages}
  {19831--19842} (\bibinfo {year} {2004})}\BibitemShut {NoStop}%
\bibitem [{\citenamefont {Schmitt}, \citenamefont {Zalabsky},\ and\
  \citenamefont {Adams}(1984)}]{Schmitt:1984}%
  \BibitemOpen
  \bibfield  {author} {\bibinfo {author} {\bibfnamefont {J.~L.}\ \bibnamefont
  {Schmitt}}, \bibinfo {author} {\bibfnamefont {R.~A.}\ \bibnamefont
  {Zalabsky}}, \ and\ \bibinfo {author} {\bibfnamefont {J.~L.}\ \bibnamefont
  {Adams}},\ }\href@noop {} {\bibfield  {journal} {\bibinfo  {journal} {J.
  Phys. Chem.}\ }\textbf {\bibinfo {volume} {79}},\ \bibinfo {pages} {4496}
  (\bibinfo {year} {1984})}\BibitemShut {NoStop}%
\bibitem [{\citenamefont {Wright}\ \emph {et~al.}(1993)\citenamefont {Wright},
  \citenamefont {Caldwell}, \citenamefont {Moxely},\ and\ \citenamefont
  {El-Shall}}]{Wright:1993}%
  \BibitemOpen
  \bibfield  {author} {\bibinfo {author} {\bibfnamefont {D.}~\bibnamefont
  {Wright}}, \bibinfo {author} {\bibfnamefont {R.}~\bibnamefont {Caldwell}},
  \bibinfo {author} {\bibfnamefont {C.}~\bibnamefont {Moxely}}, \ and\ \bibinfo
  {author} {\bibfnamefont {M.~S.}\ \bibnamefont {El-Shall}},\ }\href@noop {}
  {\bibfield  {journal} {\bibinfo  {journal} {J. Phys. Chem.}\ }\textbf
  {\bibinfo {volume} {98}},\ \bibinfo {pages} {3356} (\bibinfo {year}
  {1993})}\BibitemShut {NoStop}%
\bibitem [{\citenamefont {Viisanen}, \citenamefont {Strey},\ and\ \citenamefont
  {Reiss}(1993)}]{Viisanen:1993}%
  \BibitemOpen
  \bibfield  {author} {\bibinfo {author} {\bibfnamefont {Y.}~\bibnamefont
  {Viisanen}}, \bibinfo {author} {\bibfnamefont {R.}~\bibnamefont {Strey}}, \
  and\ \bibinfo {author} {\bibfnamefont {H.}~\bibnamefont {Reiss}},\
  }\href@noop {} {\bibfield  {journal} {\bibinfo  {journal} {J. Phys. Chem.}\
  }\textbf {\bibinfo {volume} {99}},\ \bibinfo {pages} {4680} (\bibinfo {year}
  {1993})}\BibitemShut {NoStop}%
\bibitem [{\citenamefont {Anisimov}\ \emph {et~al.}(1993)\citenamefont
  {Anisimov}, \citenamefont {Hopke}, \citenamefont {Shaimordanov},
  \citenamefont {Shandakov},\ and\ \citenamefont {Magnusson}}]{Anisimov:2001}%
  \BibitemOpen
  \bibfield  {author} {\bibinfo {author} {\bibfnamefont {M.~P.}\ \bibnamefont
  {Anisimov}}, \bibinfo {author} {\bibfnamefont {P.~K.}\ \bibnamefont {Hopke}},
  \bibinfo {author} {\bibfnamefont {I.~N.}\ \bibnamefont {Shaimordanov}},
  \bibinfo {author} {\bibfnamefont {S.~D.}\ \bibnamefont {Shandakov}}, \ and\
  \bibinfo {author} {\bibfnamefont {L.-E.}\ \bibnamefont {Magnusson}},\
  }\href@noop {} {\bibfield  {journal} {\bibinfo  {journal} {J. Phys. Chem.}\
  }\textbf {\bibinfo {volume} {115}},\ \bibinfo {pages} {810} (\bibinfo {year}
  {1993})}\BibitemShut {NoStop}%
\bibitem [{\citenamefont {Sinha}\ \emph {et~al.}(2010)\citenamefont {Sinha},
  \citenamefont {Bhabhe}, \citenamefont {Laksmono}, \citenamefont {W{\"o}lk},
  \citenamefont {Strey},\ and\ \citenamefont {Wyslouzil}}]{Sinha:2001}%
  \BibitemOpen
  \bibfield  {author} {\bibinfo {author} {\bibfnamefont {S.}~\bibnamefont
  {Sinha}}, \bibinfo {author} {\bibfnamefont {A.}~\bibnamefont {Bhabhe}},
  \bibinfo {author} {\bibfnamefont {H.}~\bibnamefont {Laksmono}}, \bibinfo
  {author} {\bibfnamefont {J.}~\bibnamefont {W{\"o}lk}}, \bibinfo {author}
  {\bibfnamefont {R.}~\bibnamefont {Strey}}, \ and\ \bibinfo {author}
  {\bibfnamefont {B.}~\bibnamefont {Wyslouzil}},\ }\href@noop {} {\bibfield
  {journal} {\bibinfo  {journal} {J. Phys. Chem.}\ }\textbf {\bibinfo {volume}
  {132}},\ \bibinfo {pages} {064304} (\bibinfo {year} {2010})}\BibitemShut
  {NoStop}%
\bibitem [{\citenamefont {Kalikmanov}(2006)}]{Kalikmanov:2006}%
  \BibitemOpen
  \bibfield  {author} {\bibinfo {author} {\bibfnamefont {V.}~\bibnamefont
  {Kalikmanov}},\ }\href@noop {} {\bibfield  {journal} {\bibinfo  {journal} {J.
  Chem. Phys.}\ }\textbf {\bibinfo {volume} {124(12)}},\ \bibinfo {pages}
  {124505} (\bibinfo {year} {2006})}\BibitemShut {NoStop}%
\bibitem [{\citenamefont {Kusaka}, \citenamefont {Wang},\ and\ \citenamefont
  {Seinfeld}(1998)}]{Kusaka:1998b}%
  \BibitemOpen
  \bibfield  {author} {\bibinfo {author} {\bibfnamefont {I.}~\bibnamefont
  {Kusaka}}, \bibinfo {author} {\bibfnamefont {Z.-G.}\ \bibnamefont {Wang}}, \
  and\ \bibinfo {author} {\bibfnamefont {J.~H.}\ \bibnamefont {Seinfeld}},\
  }\href@noop {} {\bibfield  {journal} {\bibinfo  {journal} {J. Phys. Chem.}\
  }\textbf {\bibinfo {volume} {108}},\ \bibinfo {pages} {3416} (\bibinfo {year}
  {1998})}\BibitemShut {NoStop}%
\bibitem [{\citenamefont {Oh}\ and\ \citenamefont {Zeng}(1999)}]{Oh:1999}%
  \BibitemOpen
  \bibfield  {author} {\bibinfo {author} {\bibfnamefont {K.~J.}\ \bibnamefont
  {Oh}}\ and\ \bibinfo {author} {\bibfnamefont {X.~C.}\ \bibnamefont {Zeng}},\
  }\href@noop {} {\bibfield  {journal} {\bibinfo  {journal} {J. Phys. Chem.}\
  }\textbf {\bibinfo {volume} {110}},\ \bibinfo {pages} {4471} (\bibinfo {year}
  {1999})}\BibitemShut {NoStop}%
\bibitem [{\citenamefont {Oh}\ and\ \citenamefont {Zeng}(2000)}]{Oh:2000}%
  \BibitemOpen
  \bibfield  {author} {\bibinfo {author} {\bibfnamefont {K.~J.}\ \bibnamefont
  {Oh}}\ and\ \bibinfo {author} {\bibfnamefont {X.~C.}\ \bibnamefont {Zeng}},\
  }\href@noop {} {\bibfield  {journal} {\bibinfo  {journal} {J. Phys. Chem.}\
  }\textbf {\bibinfo {volume} {112}},\ \bibinfo {pages} {294} (\bibinfo {year}
  {2000})}\BibitemShut {NoStop}%
\bibitem [{\citenamefont {Chen}\ \emph {et~al.}(2001)\citenamefont {Chen},
  \citenamefont {Siepmann}, \citenamefont {Kwang},\ and\ \citenamefont
  {Klein}}]{Chen:2001}%
  \BibitemOpen
  \bibfield  {author} {\bibinfo {author} {\bibfnamefont {B.}~\bibnamefont
  {Chen}}, \bibinfo {author} {\bibfnamefont {I.}~\bibnamefont {Siepmann}},
  \bibinfo {author} {\bibfnamefont {J.~O.}\ \bibnamefont {Kwang}}, \ and\
  \bibinfo {author} {\bibfnamefont {M.~L.}\ \bibnamefont {Klein}},\ }\href@noop
  {} {\bibfield  {journal} {\bibinfo  {journal} {J. Phys. Chem.}\ }\textbf
  {\bibinfo {volume} {115}},\ \bibinfo {pages} {10903} (\bibinfo {year}
  {2001})}\BibitemShut {NoStop}%
\bibitem [{\citenamefont {Yoo}, \citenamefont {Oh},\ and\ \citenamefont
  {Zeng}(2001)}]{Yoo:2001}%
  \BibitemOpen
  \bibfield  {author} {\bibinfo {author} {\bibfnamefont {S.}~\bibnamefont
  {Yoo}}, \bibinfo {author} {\bibfnamefont {K.~J.}\ \bibnamefont {Oh}}, \ and\
  \bibinfo {author} {\bibfnamefont {X.~C.}\ \bibnamefont {Zeng}},\ }\href@noop
  {} {\bibfield  {journal} {\bibinfo  {journal} {J. Phys. Chem.}\ }\textbf
  {\bibinfo {volume} {115}},\ \bibinfo {pages} {8518} (\bibinfo {year}
  {2001})}\BibitemShut {NoStop}%
\bibitem [{\citenamefont {Gonzalez}\ \emph {et~al.}(2014)\citenamefont
  {Gonzalez}, \citenamefont {Sanz}, \citenamefont {McBride}, \citenamefont
  {Abascal}, \citenamefont {Vegaa},\ and\ \citenamefont
  {Valeriani}}]{Gonzalez:2014}%
  \BibitemOpen
  \bibfield  {author} {\bibinfo {author} {\bibfnamefont {M.~A.}\ \bibnamefont
  {Gonzalez}}, \bibinfo {author} {\bibfnamefont {E.}~\bibnamefont {Sanz}},
  \bibinfo {author} {\bibfnamefont {C.}~\bibnamefont {McBride}}, \bibinfo
  {author} {\bibfnamefont {J.~L.~F.}\ \bibnamefont {Abascal}}, \bibinfo
  {author} {\bibfnamefont {C.}~\bibnamefont {Vegaa}}, \ and\ \bibinfo {author}
  {\bibfnamefont {C.}~\bibnamefont {Valeriani}},\ }\href@noop {} {\bibfield
  {journal} {\bibinfo  {journal} {J. Phys. Chem.}\ }\textbf {\bibinfo {volume}
  {16}},\ \bibinfo {pages} {24913--24919} (\bibinfo {year} {2014})}\BibitemShut
  {NoStop}%
\bibitem [{\citenamefont {ten Wolde}\ and\ \citenamefont
  {Frenkel}(1998)}]{Wolde:1998}%
  \BibitemOpen
  \bibfield  {author} {\bibinfo {author} {\bibfnamefont {P.~R.}\ \bibnamefont
  {ten Wolde}}\ and\ \bibinfo {author} {\bibfnamefont {D.}~\bibnamefont
  {Frenkel}},\ }\href@noop {} {\bibfield  {journal} {\bibinfo  {journal} {J.
  Phys. Chem.}\ }\textbf {\bibinfo {volume} {109}},\ \bibinfo {pages} {9901}
  (\bibinfo {year} {1998})}\BibitemShut {NoStop}%
\bibitem [{\citenamefont {Schweizer}\ and\ \citenamefont
  {Sagis}(2014)}]{Schweizer:2014}%
  \BibitemOpen
  \bibfield  {author} {\bibinfo {author} {\bibfnamefont {M.}~\bibnamefont
  {Schweizer}}\ and\ \bibinfo {author} {\bibfnamefont {L.}~\bibnamefont
  {Sagis}},\ }\href@noop {} {\bibfield  {journal} {\bibinfo  {journal} {J.
  Chem. Phys.}\ }\textbf {\bibinfo {volume} {141}},\ \bibinfo {pages} {224102}
  (\bibinfo {year} {2014})}\BibitemShut {NoStop}%
\bibitem [{\citenamefont {Feder}, \citenamefont {Lothe},\ and\ \citenamefont
  {Pound}(1966)}]{Feder:1966}%
  \BibitemOpen
  \bibfield  {author} {\bibinfo {author} {\bibfnamefont {J.}~\bibnamefont
  {Feder}}, \bibinfo {author} {\bibfnamefont {J.}~\bibnamefont {Lothe}}, \ and\
  \bibinfo {author} {\bibfnamefont {G.~M.}\ \bibnamefont {Pound}},\ }\href@noop
  {} {\bibfield  {journal} {\bibinfo  {journal} {Adv. Phys.}\ }\textbf
  {\bibinfo {volume} {15}},\ \bibinfo {pages} {111} (\bibinfo {year}
  {1966})}\BibitemShut {NoStop}%
\bibitem [{\citenamefont {Barrett}(2008)}]{Barrett:2008}%
  \BibitemOpen
  \bibfield  {author} {\bibinfo {author} {\bibfnamefont {J.~C.}\ \bibnamefont
  {Barrett}},\ }\href@noop {} {\bibfield  {journal} {\bibinfo  {journal} {J.
  Chem. Phys.}\ }\textbf {\bibinfo {volume} {128}},\ \bibinfo {pages} {164519}
  (\bibinfo {year} {2008})}\BibitemShut {NoStop}%
\bibitem [{\citenamefont {Barrett}(1994)}]{Barrett:1994}%
  \BibitemOpen
  \bibfield  {author} {\bibinfo {author} {\bibfnamefont {J.~C.}\ \bibnamefont
  {Barrett}},\ }\href@noop {} {\bibfield  {journal} {\bibinfo  {journal} {J.
  Phys. A: Math. Gen.}\ }\textbf {\bibinfo {volume} {27}},\ \bibinfo {pages}
  {5053} (\bibinfo {year} {1994})}\BibitemShut {NoStop}%
\bibitem [{\citenamefont {Wyslouzil}\ and\ \citenamefont
  {Seinfeld}(1992)}]{Wyslouzil:1992}%
  \BibitemOpen
  \bibfield  {author} {\bibinfo {author} {\bibfnamefont {B.~E.}\ \bibnamefont
  {Wyslouzil}}\ and\ \bibinfo {author} {\bibfnamefont {J.~H.}\ \bibnamefont
  {Seinfeld}},\ }\href@noop {} {\bibfield  {journal} {\bibinfo  {journal} {J.
  Chem. Phys.}\ }\textbf {\bibinfo {volume} {97}},\ \bibinfo {pages} {2661}
  (\bibinfo {year} {1992})}\BibitemShut {NoStop}%
\bibitem [{\citenamefont {Lothe}\ and\ \citenamefont
  {Pound}(1962)}]{Lothe:1962}%
  \BibitemOpen
  \bibfield  {author} {\bibinfo {author} {\bibfnamefont {J.}~\bibnamefont
  {Lothe}}\ and\ \bibinfo {author} {\bibfnamefont {G.~M.}\ \bibnamefont
  {Pound}},\ }\href@noop {} {\bibfield  {journal} {\bibinfo  {journal} {J.
  Chem. Phys.}\ }\textbf {\bibinfo {volume} {36}},\ \bibinfo {pages} {2080}
  (\bibinfo {year} {1962})}\BibitemShut {NoStop}%
\bibitem [{\citenamefont {Reiss}, \citenamefont {Kegel},\ and\ \citenamefont
  {Katz}(1998)}]{Reiss:1998}%
  \BibitemOpen
  \bibfield  {author} {\bibinfo {author} {\bibfnamefont {H.}~\bibnamefont
  {Reiss}}, \bibinfo {author} {\bibfnamefont {W.~K.}\ \bibnamefont {Kegel}}, \
  and\ \bibinfo {author} {\bibfnamefont {J.~L.}\ \bibnamefont {Katz}},\
  }\href@noop {} {\bibfield  {journal} {\bibinfo  {journal} {Phys. Chem. A}\
  }\textbf {\bibinfo {volume} {102}},\ \bibinfo {pages} {8548} (\bibinfo {year}
  {1998})}\BibitemShut {NoStop}%
\bibitem [{\citenamefont {Reiss}, \citenamefont {Kegel},\ and\ \citenamefont
  {Katz}(1997)}]{Reiss:1997}%
  \BibitemOpen
  \bibfield  {author} {\bibinfo {author} {\bibfnamefont {H.}~\bibnamefont
  {Reiss}}, \bibinfo {author} {\bibfnamefont {W.~K.}\ \bibnamefont {Kegel}}, \
  and\ \bibinfo {author} {\bibfnamefont {J.~L.}\ \bibnamefont {Katz}},\
  }\href@noop {} {\bibfield  {journal} {\bibinfo  {journal} {Phys. Rev. Lett.}\
  }\textbf {\bibinfo {volume} {78}},\ \bibinfo {pages} {4506} (\bibinfo {year}
  {1997})}\BibitemShut {NoStop}%
\bibitem [{\citenamefont {Ruth}, \citenamefont {Hirth},\ and\ \citenamefont
  {Pound}(1988)}]{Ruth:1988}%
  \BibitemOpen
  \bibfield  {author} {\bibinfo {author} {\bibfnamefont {V.}~\bibnamefont
  {Ruth}}, \bibinfo {author} {\bibfnamefont {J.~P.}\ \bibnamefont {Hirth}}, \
  and\ \bibinfo {author} {\bibfnamefont {G.~M.}\ \bibnamefont {Pound}},\
  }\href@noop {} {\bibfield  {journal} {\bibinfo  {journal} {J. Chem. Phys.}\
  }\textbf {\bibinfo {volume} {88}},\ \bibinfo {pages} {7079} (\bibinfo {year}
  {1988})}\BibitemShut {NoStop}%
\bibitem [{\citenamefont {Reiss}, \citenamefont {Katz},\ and\ \citenamefont
  {Cohe}(1968)}]{Reiss:1967b}%
  \BibitemOpen
  \bibfield  {author} {\bibinfo {author} {\bibfnamefont {H.}~\bibnamefont
  {Reiss}}, \bibinfo {author} {\bibfnamefont {J.~L.}\ \bibnamefont {Katz}}, \
  and\ \bibinfo {author} {\bibfnamefont {E.~R.}\ \bibnamefont {Cohe}},\
  }\href@noop {} {\bibfield  {journal} {\bibinfo  {journal} {J. Chem. Phys.}\
  }\textbf {\bibinfo {volume} {48}},\ \bibinfo {pages} {5553} (\bibinfo {year}
  {1968})}\BibitemShut {NoStop}%
\bibitem [{\citenamefont {Reiss}\ and\ \citenamefont
  {Katz}(1967)}]{Reiss:1967a}%
  \BibitemOpen
  \bibfield  {author} {\bibinfo {author} {\bibfnamefont {H.}~\bibnamefont
  {Reiss}}\ and\ \bibinfo {author} {\bibfnamefont {J.~L.}\ \bibnamefont
  {Katz}},\ }\href@noop {} {\bibfield  {journal} {\bibinfo  {journal} {J. Chem.
  Phys.}\ }\textbf {\bibinfo {volume} {46}},\ \bibinfo {pages} {2496} (\bibinfo
  {year} {1967})}\BibitemShut {NoStop}%
\bibitem [{\citenamefont {Reguera}\ and\ \citenamefont
  {Rubi}(2001)}]{Reguera:2001q}%
  \BibitemOpen
  \bibfield  {author} {\bibinfo {author} {\bibfnamefont {D.}~\bibnamefont
  {Reguera}}\ and\ \bibinfo {author} {\bibfnamefont {J.~M.}\ \bibnamefont
  {Rubi}},\ }\href@noop {} {\bibfield  {journal} {\bibinfo  {journal} {J. Chem.
  Phys.}\ }\textbf {\bibinfo {volume} {115}},\ \bibinfo {pages} {7100}
  (\bibinfo {year} {2001})}\BibitemShut {NoStop}%
\bibitem [{\citenamefont {Schaaf}\ \emph {et~al.}(1999)\citenamefont {Schaaf},
  \citenamefont {Senger}, \citenamefont {Voegel},\ and\ \citenamefont
  {Reiss}}]{Schaaf:1999oparam}%
  \BibitemOpen
  \bibfield  {author} {\bibinfo {author} {\bibfnamefont {P.}~\bibnamefont
  {Schaaf}}, \bibinfo {author} {\bibfnamefont {B.}~\bibnamefont {Senger}},
  \bibinfo {author} {\bibfnamefont {J.-C.}\ \bibnamefont {Voegel}}, \ and\
  \bibinfo {author} {\bibfnamefont {H.}~\bibnamefont {Reiss}},\ }\href@noop {}
  {\bibfield  {journal} {\bibinfo  {journal} {Phys. Rev. E}\ }\textbf {\bibinfo
  {volume} {60}},\ \bibinfo {pages} {771} (\bibinfo {year} {1999})}\BibitemShut
  {NoStop}%
\bibitem [{\citenamefont {Kusaka}(2003)}]{Kusaka:2003oparam}%
  \BibitemOpen
  \bibfield  {author} {\bibinfo {author} {\bibfnamefont {I.}~\bibnamefont
  {Kusaka}},\ }\href@noop {} {\bibfield  {journal} {\bibinfo  {journal} {J.
  Chem. Phys.}\ }\textbf {\bibinfo {volume} {119}},\ \bibinfo {pages} {3820}
  (\bibinfo {year} {2003})}\BibitemShut {NoStop}%
\bibitem [{\citenamefont {Kusaka}\ and\ \citenamefont
  {Oxtoby}(1999)}]{Kusaka:1999oparam}%
  \BibitemOpen
  \bibfield  {author} {\bibinfo {author} {\bibfnamefont {I.}~\bibnamefont
  {Kusaka}}\ and\ \bibinfo {author} {\bibfnamefont {D.~W.}\ \bibnamefont
  {Oxtoby}},\ }\href@noop {} {\bibfield  {journal} {\bibinfo  {journal} {J.
  Chem. Phys.}\ }\textbf {\bibinfo {volume} {110}},\ \bibinfo {pages} {5249}
  (\bibinfo {year} {1999})}\BibitemShut {NoStop}%
\bibitem [{\citenamefont {\"Ottinger}(2007)}]{HCO:POT}%
  \BibitemOpen
  \bibfield  {author} {\bibinfo {author} {\bibfnamefont {H.~C.}\ \bibnamefont
  {\"Ottinger}},\ }\href@noop {} {\bibfield  {journal} {\bibinfo  {journal}
  {MRS Bull.}\ }\textbf {\bibinfo {volume} {32}},\ \bibinfo {pages} {936--940}
  (\bibinfo {year} {2007})}\BibitemShut {NoStop}%
\bibitem [{\citenamefont {\"Ottinger}(2005)}]{HCO}%
  \BibitemOpen
  \bibfield  {author} {\bibinfo {author} {\bibfnamefont {H.~C.}\ \bibnamefont
  {\"Ottinger}},\ }\href@noop {} {\emph {\bibinfo {title} {Beyond Equilibrium
  Thermodynamics}}}\ (\bibinfo  {publisher} {Wiley Interscience},\ \bibinfo
  {year} {2005})\BibitemShut {NoStop}%
\bibitem [{\citenamefont {Lundrigan}\ and\ \citenamefont
  {Saika-Voivod}(2009)}]{Lundrigan:2009}%
  \BibitemOpen
  \bibfield  {author} {\bibinfo {author} {\bibfnamefont {S.~E.~M.}\
  \bibnamefont {Lundrigan}}\ and\ \bibinfo {author} {\bibfnamefont
  {I.}~\bibnamefont {Saika-Voivod}},\ }\href@noop {} {\bibfield  {journal}
  {\bibinfo  {journal} {J. Chem. Phys.}\ }\textbf {\bibinfo {volume} {131}},\
  \bibinfo {pages} {104503} (\bibinfo {year} {2009})}\BibitemShut {NoStop}%
\bibitem [{\citenamefont {Auer}\ and\ \citenamefont
  {Frenkel}(2004)}]{Auer:2004}%
  \BibitemOpen
  \bibfield  {author} {\bibinfo {author} {\bibfnamefont {S.}~\bibnamefont
  {Auer}}\ and\ \bibinfo {author} {\bibfnamefont {D.}~\bibnamefont {Frenkel}},\
  }\href@noop {} {\bibfield  {journal} {\bibinfo  {journal} {J. Chem. Phys.}\
  }\textbf {\bibinfo {volume} {120}},\ \bibinfo {pages} {3015} (\bibinfo {year}
  {2004})}\BibitemShut {NoStop}%
\bibitem [{\citenamefont {McGraw}\ and\ \citenamefont {A}(1995)}]{McGraw:1995}%
  \BibitemOpen
  \bibfield  {author} {\bibinfo {author} {\bibfnamefont {R.}~\bibnamefont
  {McGraw}}\ and\ \bibinfo {author} {\bibfnamefont {L.~R.}\ \bibnamefont {A}},\
  }\href@noop {} {\bibfield  {journal} {\bibinfo  {journal} {J. Chem. Phys.}\
  }\textbf {\bibinfo {volume} {102}},\ \bibinfo {pages} {8983} (\bibinfo {year}
  {1995})}\BibitemShut {NoStop}%
\bibitem [{\citenamefont {Wedekind}, \citenamefont {D},\ and\ \citenamefont
  {Strey}(2007)}]{Wedekind:2007aa}%
  \BibitemOpen
  \bibfield  {author} {\bibinfo {author} {\bibfnamefont {J.}~\bibnamefont
  {Wedekind}}, \bibinfo {author} {\bibfnamefont {R.}~\bibnamefont {D}}, \ and\
  \bibinfo {author} {\bibfnamefont {R.}~\bibnamefont {Strey}},\ }\href@noop {}
  {\bibfield  {journal} {\bibinfo  {journal} {J. Chem. Phys.}\ }\textbf
  {\bibinfo {volume} {127}},\ \bibinfo {pages} {064501} (\bibinfo {year}
  {2007})}\BibitemShut {NoStop}%
\bibitem [{\citenamefont {Ang\'{e}lil}\ \emph {et~al.}(2014)\citenamefont
  {Ang\'{e}lil}, \citenamefont {Diemand}, \citenamefont {Tanaka},\ and\
  \citenamefont {Tanaka}}]{Diemand:2014ads}%
  \BibitemOpen
  \bibfield  {author} {\bibinfo {author} {\bibfnamefont {R.}~\bibnamefont
  {Ang\'{e}lil}}, \bibinfo {author} {\bibfnamefont {J.}~\bibnamefont
  {Diemand}}, \bibinfo {author} {\bibfnamefont {K.~K.}\ \bibnamefont {Tanaka}},
  \ and\ \bibinfo {author} {\bibfnamefont {H.}~\bibnamefont {Tanaka}},\
  }\href@noop {} {\bibfield  {journal} {\bibinfo  {journal} {J. Chem. Phys.}\
  }\textbf {\bibinfo {volume} {140}},\ \bibinfo {pages} {074303} (\bibinfo
  {year} {2014})}\BibitemShut {NoStop}%
\bibitem [{\citenamefont {Wedekind}\ and\ \citenamefont
  {Reguera}(2007)}]{Wedekind:2007adee}%
  \BibitemOpen
  \bibfield  {author} {\bibinfo {author} {\bibfnamefont {J.}~\bibnamefont
  {Wedekind}}\ and\ \bibinfo {author} {\bibfnamefont {D.}~\bibnamefont
  {Reguera}},\ }\href@noop {} {\bibfield  {journal} {\bibinfo  {journal} {J.
  Chem. Phys.}\ }\textbf {\bibinfo {volume} {127(15)}},\ \bibinfo {pages}
  {154516} (\bibinfo {year} {2007})}\BibitemShut {NoStop}%
\bibitem [{\citenamefont {Reguera}\ and\ \citenamefont
  {Rubi}(2003{\natexlab{a}})}]{Reguera:2003tt}%
  \BibitemOpen
  \bibfield  {author} {\bibinfo {author} {\bibfnamefont {D.}~\bibnamefont
  {Reguera}}\ and\ \bibinfo {author} {\bibfnamefont {J.~M.}\ \bibnamefont
  {Rubi}},\ }\href@noop {} {\bibfield  {journal} {\bibinfo  {journal} {J. Chem.
  Phys.}\ }\textbf {\bibinfo {volume} {119}},\ \bibinfo {pages} {9888}
  (\bibinfo {year} {2003}{\natexlab{a}})}\BibitemShut {NoStop}%
\bibitem [{\citenamefont {Reguera}\ and\ \citenamefont
  {Rubi}(2003{\natexlab{b}})}]{Reguera:2003ttt}%
  \BibitemOpen
  \bibfield  {author} {\bibinfo {author} {\bibfnamefont {D.}~\bibnamefont
  {Reguera}}\ and\ \bibinfo {author} {\bibfnamefont {J.~M.}\ \bibnamefont
  {Rubi}},\ }\href@noop {} {\bibfield  {journal} {\bibinfo  {journal} {J. Chem.
  Phys.}\ }\textbf {\bibinfo {volume} {119}},\ \bibinfo {pages} {9877}
  (\bibinfo {year} {2003}{\natexlab{b}})}\BibitemShut {NoStop}%
\bibitem [{\citenamefont {Trinkaus}(1983)}]{Trinkaus:1983}%
  \BibitemOpen
  \bibfield  {author} {\bibinfo {author} {\bibfnamefont {H.}~\bibnamefont
  {Trinkaus}},\ }\href@noop {} {\bibfield  {journal} {\bibinfo  {journal}
  {Phys. Rev. B}\ }\textbf {\bibinfo {volume} {27}},\ \bibinfo {pages} {7372}
  (\bibinfo {year} {1983})}\BibitemShut {NoStop}%
\bibitem [{\citenamefont {Tanaka}\ \emph {et~al.}(2014)\citenamefont {Tanaka},
  \citenamefont {Diemand}, \citenamefont {Ang\'{e}lil},\ and\ \citenamefont
  {Tanaka}}]{Tanaka:2014scaling}%
  \BibitemOpen
  \bibfield  {author} {\bibinfo {author} {\bibfnamefont {K.~K.}\ \bibnamefont
  {Tanaka}}, \bibinfo {author} {\bibfnamefont {J.}~\bibnamefont {Diemand}},
  \bibinfo {author} {\bibfnamefont {R.}~\bibnamefont {Ang\'{e}lil}}, \ and\
  \bibinfo {author} {\bibfnamefont {H.}~\bibnamefont {Tanaka}},\ }\href@noop {}
  {\bibfield  {journal} {\bibinfo  {journal} {J. Chem. Phys.}\ }\textbf
  {\bibinfo {volume} {140}},\ \bibinfo {pages} {194310} (\bibinfo {year}
  {2014})}\BibitemShut {NoStop}%
\end{thebibliography}%

\section{Appendix}

\subsection{Truncated Entropy Splitting Eq.~(\ref{AMP_ENTR_SPLIT})}
\label{sec_entr_splitt}

Without restriction of generality, we can assume $\ptot = 0$ emphasizing Galilean invariance. When the environment is sufficiently large, the variables $\xtot$ reach their thermodynamic limit and by the equivalence of ensembles we then have $  S(\xtot, \xc) = \log(Z(\xtot, \xc) )$, where
\begin{equation}
\begin{split} 
Z(\xtot, \xc) &=  \int_{\Vtot} d\Gamma \  \delta(\Pi_{\xc} - \xc ) e^{-\frac{\Pi_{\etot}}{2T}  } \\&
 	        = e^{-\frac{\pc^2}{2\mc T}} Z(\xtot, \mc, \pc = 0, \mc).
\end{split}
\end{equation} 
The second line follows by straight forward calculation. This can be related to the truncated partition function $Z(\xtot, \mc, \uc)$ which finally defines the truncated entropy $S(\xtot, \mc, \uc)$ by means of
\begin{equation}
\begin{split} 
 Z(\xtot, \mc, \uc)  &= \int d^3\pc \ Z(\xtot, \xc) \\&
 	 	   = \left(\frac{\mc}{ T} \right)^{3/2} Z(\xtot, \mc, \pc = 0, \mc) \\&
 		   =  \left(\frac{\mc}{ T} \right)^{3/2}  e^{\frac{\pc^2}{2\mc T}} Z(\xtot, \xc).
\end{split}
\end{equation} 
The required relationship (\ref{AMP_ENTR_SPLIT}) than follows by exploiting Galilean invariance again to shift to the general case of arbitrary $\ptot $.

\subsection{Detailed Derivation of Evolution Equation}
\label{sec_evolution_eqn_derivation}

We use the GENERIC projection operator technique\cite{HCO:POT,HCO} in order to find the fundamental time evolution of the distribution function $f$ of the nucleus of interest. To this extend a projection operator $\Pcal$ is introduced that separates the macroscopically relevant contribution of an arbitrary observable $\Pi_{A}$ on the phase space $\bar{\Gamma}$ from the irrelevant ones to be eliminated - this corresponds to the separation of slow, relevant from fast, irrelevant degrees of freedom when switching from the finer-level of description involving all $N$ atoms to the coarse-grained level characterized by $(\xtot, f)$. Following the standard definition, we define the projection onto the slow variables 
\begin{equation}
\begin{split} 
\Pcal \Pi_A =& \langle \Pi_A \rangle_{\rho(\xtot, \lambdac)} \\&
+ \left(\Pi_{\xtot} - \xtot \right) \cdot \frac{\delta}{\delta \xtot} \langle \Pi_A \rangle_{\rho(\xtot, \lambdac)} \\&
+ \int d\xc \left(\Pi_{f(\xc)} - f(\xc) \right)  \frac{\delta}{\delta f(\xc)}\langle \Pi_A \rangle_{\rho(\xtot, \lambdac)}. 
\end{split}
\end{equation} 
Using
\begin{equation}
\begin{split} 
\rho_{\xtot, \lambdac} \Pi_{f(\xc)} = f(\xc) \rho_{\xtot, \xc},
\label{aux_prop1}
\end{split}
\end{equation} 
where 
\begin{equation}
\begin{split} 
\rho_{\xtot, \xc} \propto \delta(\Pi_{\xtot}-\xtot) \delta (\Pi_{\xc} -\xc),
\label{aux_prop1}
\end{split}
\end{equation} 
is the fully microcanonical ensemble, the projection operator can be expressed alternatively as
\begin{equation}
\begin{split} 
\Pcal \Pi_A =& \langle \Pi_A \rangle_{(\xtot, \xc = \Pi_{\xc})} \\& 
+ \left(\Pi_{\xtot} - \xtot \right) \cdot \frac{\delta}{\delta \xtot} \int d\xc  f(\xc) \langle \Pi_A \rangle_{(\xtot, \xc)}. 
\end{split}
\end{equation} 
The first part projects onto the nucleus variables $\Pi_{\xc}$ while the second part onto the system state $\xtot$. Complementary, we define $\Qcal = 1 - \Pcal$ as the projection onto the fast variables. Denoting by $\Lcal$ the Liouville operator on $\bar{\Gamma}$, we notice that $\Lcal \Pi_{\xtot} = 0$. The fact that $\xtot$ are conserved quantites also implies on the macroscale $d\xtot/dt = 0$. The projection operator technique then implies\cite{HCO:POT,HCO} the time-evolution for $f$, 
\begin{equation}
\begin{split} 
\frac{\partial f(\xc)}{\partial t} =& \langle \Lcal  \Pi_{f(\xc)}  \rangle_{\rho(\xtot, \lambdac)} \\& 
+ \int d\xc' M_{f(\xc),f(\xc')} \frac{\delta S(\xtot, f)}{\delta f(\xc')},
\label{eq_Pif_time_evo}
\end{split}
\end{equation} 
where
\begin{equation}
\begin{split} 
M_{f(\xc),f(\xc')} = \frac{1}{\kb} \int_{0}^{ \tau_{\text{GK}}} du \langle \dot{\Pi}_{f(\xc')} G(u)  \dot{\Pi}_{f(\xc)} \rangle_{\rho(\xtot, \lambdac)}.
\end{split}
\end{equation} 
Here, $\dot{\Pi}_{f(\xc)} =\Qcal\Lcal \Pi_{f(\xc)}$ is the fast time-evolution of $\Pi_f$ and $G(u) = e^{\Qcal \Lcal \Qcal u}$ is the time-evolution operator. the time evolution equation for $f$, Eq.~(\ref{eq_Pif_time_evo}), splits into a reversible part and an irreversible part involving the Green-Kubo coefficients $M_{f(\xc),f(\xc')}$ and driven by the total entropy $S(\xtot, f)$ of the system. The time-scale $ \tau_{\text{GK}}$ is an intermediate time-scale between microscopic collisions and the mesoscopic time-scale on which the nucleation process is described in a coarse-grained way.

Using the auxiliary properties
\begin{equation}
\begin{split} 
\Lcal {\Pi}_{f(\xc)} = - \frac{\delta \Pi_{f(\xc)}}{\delta \xc} \cdot \Lcal \Pi_{\xc},
\end{split}
\end{equation} 
and (\ref{aux_prop1}) the reversible part of the time-evolution can be cast into
\begin{equation}
\begin{split} 
\left. \frac{\partial f(\xc)}{\partial t}\right|_{\text{rev}} &=
\langle \Lcal  \Pi_{f(\xc)}  \rangle_{\rho(\xtot, \lambdac)}  \\&
= - \frac{\delta}{\delta \xc} f(\xc) \langle   \Lcal \Pi_{\xc} \rangle_{\rho_{\xtot, \xc}}.
\end{split}
\end{equation} 
We now turn our attention back to the projection operator $\Pcal$. This operator, or its counterpart $\Qcal$ act on $\Lcal {\Pi}_{\xc}$ in the evolution equation. Therefore, we must understand the object $\langle   \Lcal \Pi_{\xc} \rangle_{\rho_{\xtot, \xc}}$ which also enters the reversible part of the evolution equation. For homogeneous nucleation this reversible part is known to vanish, so that we already expect 
\begin{equation}
\begin{split} 
\langle   \Lcal \Pi_{\xc} \rangle_{\rho_{\xtot, \xc}} = 0
\label{abs_rev_part}
\end{split}
\end{equation}  
This implies $\Pcal \Lcal \Pi_{\xc} = 0$ and $\Qcal  \Lcal \Pi_{\xc}  =  \Lcal \Pi_{\xc} $. The argument for Eq.~(\ref{abs_rev_part}) can be made more formal by recognizing that the average over $ \Lcal \Pi_{\xc}$ must be taken over all states compatible with the microcanonical ensemble $\rho_{\xtot, \xc}$, that is over all states for which we encounter a nucleus with properties $\xc$. But if a certain atomistic configuration $C = (\bm{r}_1, ..., \bm{r}_N,  \bm{p}_1, ... \bm{p}_N)$ contains a nucleus with properties $\xc$, then also the mirror image $C' = (-\bm{r}_1, ..., -\bm{r}_N,  \bm{p}_1, ... \bm{p}_N)$ obeys this requirement. It is important that the momentum variables must not be mirror imaged, since otherwise the momentum $\Pi_{\pc}$ of $C$ and $C'$ do not match and therefore $C'$ would not fall into the class $\xc$. In $C$ all particles that would be about to join the nucleus would be about to leave it in $C'$ and vice verca. In the same manner one can argue that in $C$ a momentum influx is transformed to a momentum outflux in $C'$ and vice verca. The same argumentation holds with the energy flux and consequently we have $\Lcal \Pi_{\xc}(C) = - \Lcal \Pi_{\xc}(C')$ and therefore the validity of Eq.~(\ref{abs_rev_part}). 

In general, the irreversible part of the evolution equation can be cast into the form\cite{HCO:POT,HCO}
\begin{equation}
\begin{split} 
\left. \frac{\partial f(\xc)}{\partial t}\right|_{\text{irr}} &=
 \int d\xc' M_{f(\xc),f(\xc')} \frac{\delta S(\xtot, f)}{\delta f(\xc')} \\&
=
- \frac{\delta}{\delta \xc} \cdot f(\xc) M(\xc) \cdot \frac{\delta S(\xtot, f) }{\delta \xc},
\label{COMPL_GEN_EQ}
\end{split}
\end{equation} 
where 
\begin{equation}
\begin{split} 
M(\xc) &= \frac{1}{\kb} \int_{0}^{ \tau_{\text{GK}}} du \langle \Lcal \Pi_{\xc} G(u)\Lcal \Pi_{\xc} \rangle_{\rho(\xtot, \xtot)} \\&
=\frac{1}{2\kb  \tau_{\text{GK}}} \langle \Delta_{ \tau_{\text{GK}}}\Pi_{\xc} \Delta_{ \tau_{\text{GK}}}\Pi_{\xc} \rangle_{\rho(\xtot, \xtot)}.
\end{split}
\end{equation} 
We hence derived the fundamental nucleus evolution equation based on microscopic considerations.

\end{document}